\documentclass[a4paper,11pt,DIV=12,abstract=true,numbers=noenddot,titlepage=false,twocolumn=false,draft=false]{scrartcl}
\pdfoutput=1 

\makeatletter
\DeclareOldFontCommand{\rm}{\normalfont\rmfamily}{\mathrm}
\DeclareOldFontCommand{\sf}{\normalfont\sffamily}{\mathsf}
\DeclareOldFontCommand{\tt}{\normalfont\ttfamily}{\mathtt}
\DeclareOldFontCommand{\bf}{\normalfont\bfseries}{\mathbf}
\DeclareOldFontCommand{\it}{\normalfont\itshape}{\mathit}
\DeclareOldFontCommand{\sl}{\normalfont\slshape}{\@nomath\sl}

\usepackage[usenames,dvipsnames]{color}
  \definecolor{hgreen}{rgb}{0,.3,0}
  \definecolor{hred}{rgb}{.3,0,0}
  \definecolor{orange}{rgb}{1,0.5,0}
  \definecolor{hblue}{rgb}{0,0,.3}
  \definecolor{LightGray}{gray}{0.95}
  \definecolor{gray}{gray}{0.6}
\usepackage[
	    colorlinks=true,
	    linkcolor=hblue,
	    citecolor=hgreen,
	    filecolor=hblue,
	    urlcolor=hred
	    ]{hyperref}
\usepackage{mathrsfs}
\usepackage[intlimits]{amsmath}
\usepackage{amssymb}
\usepackage{slashed}
\usepackage[titletoc,title]{appendix}
\usepackage[affil-it]{authblk}
\usepackage{libertine}
\usepackage[numbers,sort&compress]{natbib}
\usepackage{graphicx}
\usepackage{layouts}
\usepackage{fancyhdr}
\usepackage{microtype}

\allowdisplaybreaks  
\setcapindent{1em}
\setkomafont{captionlabel}{\bfseries}
\setkomafont{caption}{\itshape}

\newcommand{\Lag}{\mathscr{L}}
\newcommand{\muew}{\mu_{\text{ew}}}
\newcommand{\muhad}{\mu_{\text{had}}}

\newcommand{\qp}{{q^{\prime}}}

\newcommand{\xhw}{x_{hW}}
\newcommand{\xth}{x_{th}}

\newcommand{\xtz}{x_{tZ}}

\newcommand{\xwh}{x_{Wh}}
\newcommand{\xhz}{x_{hZ}}
\newcommand{\xzh}{x_{Zh}}

\newcommand{\xfph}{x_{f'h}}

\newcommand{\xfpz}{x_{f' Z}}

\definecolor{Blu}{rgb}{0.,0.,1.}
\definecolor{Red}{rgb}{1.,0.,0.}
\definecolor{Darkgreen}{rgb}{0,0.5,0.}
\definecolor{Purple}{rgb}{0.5,0.,0.5}

\fancypagestyle{preprint}{
   \fancyhead[LE,RO]{DO-TH 22/09}
   
   }

\begin{document}

\title{\boldmath Global Constraints on Yukawa Operators in the Standard Model Effective Theory}
  
\date{\today}
\renewcommand\Authands{, }
\author[a]{Joachim~Brod%
        \thanks{\texttt{joachim.brod@uc.edu}}}
\author[b]{Jonathan~M.~Cornell%
        \thanks{\texttt{jonathancornell@weber.edu}}}
\author[c]{Dimitrios~Skodras%
        \thanks{\texttt{dimitrios.skodras@tu-dortmund.de}}}
\author[c]{Emmanuel~Stamou%
        \thanks{\texttt{emmanuel.stamou@tu-dortmund.de}}}

\affil[a]{{\large Department of Physics, University of Cincinnati, Cincinnati, OH 45221, USA}}
\affil[b]{{\large Department of Physics, Weber State University, Ogden, UT 84408, USA}}
\affil[c]{{\large Fakult\"at f\"ur Physik, TU Dortmund, D-44221 Dortmund, Germany}}

\maketitle
\thispagestyle{preprint}

\begin{abstract}
\normalsize CP-violating contributions to Higgs--fermion couplings are
absent in the standard model of particle physics (SM), but are
motivated by models of electroweak baryogenesis. Here, we employ the
framework of the SM effective theory (SMEFT) to parameterise
deviations from SM Yukawa couplings. We present the leading
contributions of the relevant operators to the fermionic electric
dipole moments (EDMs). We obtain constraints on the SMEFT Wilson
coefficients from the combination of LHC data and experimental bounds
on the electron, neutron, and mercury EDMs, and for the first time, we
perform a combined fit to LHC and EDM data allowing the presence of
CP-violating contributions from several fermion species
simultaneously. Among other results, we find non-trivial correlations
between EDM and LHC constraints even in the multi-parameter scans, for
instance, when floating the CP-even and CP-odd couplings to all
third-generation fermions.
\end{abstract}

%\pdfbookmark[1]{Table of Contents}{tableofcontents}
\setcounter{page}{1}
%\tableofcontents
%    \the\textwidth\\
%    \printinunitsof{cm}\prntlen{\textwidth}\\
%    \printinunitsof{in}\prntlen{\textwidth}

\section{Introduction\label{sec:introduction}}

Charge-Parity (CP) violating contributions to Higgs--fermion couplings
are a well-motivated possibility of physics beyond the standard model
(SM) that might help address the problem of baryogenesis with new dynamics at the 
electroweak scale. As is well known, any such contributions are strongly
constrained by null measurements of electric dipole moments (EDMs),
and less strongly by Higgs production and decay data from
colliders. It is also well known that the presence of several
CP-violating phases can lead to cancellations and thus weaker
bounds~\cite{Huber:2006ri}. The interplay of EDM and collider
constraints, in the presence of several CP-violating Higgs couplings 
to fermions simultaneously, is less well-known. In Ref.~\cite{Chien:2015xha}
constraints in the presence of two CP-violating parameters have been
studied in detail. In this work, we scan over up to six different
parameters.

Frequently, model-independent bounds on such interactions have
obtained in the so-called ``$\kappa$ framework'' by rescaling the SM
Yukawa coupling by an overall factor and a complex phase. In
Ref.~\cite{Brod:2013cka}, bounds on these factors have been obtained
by studying contributions to EDMs through Barr--Zee diagrams with
internal fermion loops. In Refs.~\cite{Altmannshofer:2015qra,
  Brod:2018lbf} it was shown by explicit calculation within the
$\kappa$ framework that bosonic two-loop diagrams have a large impact
on the EDM bounds for light quarks. However, it was later pointed out
that the way these bosonic contributions had been been computed lead
to a gauge dependent result~\cite{Altmannshofer:2020shb}. This is
related to the fact the naive implementation of the $\kappa$ framework, in 
which only the dimension-four Higgs couplings are modified, is not a
consistent quantum field theory. A consistent alternative that
resembles the $\kappa$ framework most closely would be to use Higgs
effective theory (HEFT)~\cite{Feruglio:1992wf}. In HEFT, the
dimension-four couplings are complemented by the necessary higher
dimension interactions to facilitate gauge-independent results.

Another consistent way of obtaining model-independent bounds for
CP-violating Higgs-fermion interactions is the use of the SM effective
field theory (SMEFT)~\cite{Buchmuller:1985jz}; see
Refs.~\cite{Chien:2015xha, Cirigliano:2016njn, Cirigliano:2016nyn,
  Kley:2021yhn} for recent work in this direction. We take the same
approach in this article.

For the purpose of this work, we assume that the contributions to the
EDMs of the SM fermions arise from physics at a scale $\Lambda$ that
is sufficiently higher than the electroweak scale. This permits us to
parameterize deviations from the SM in terms of operators in SMEFT. We
will assume that any of the operator of the form $(H^\dagger H)
\bar{F}_L f_R H$ may have non-zero coefficients.
Here, $H$ is the
complex Higgs doublet field, $F_L$ a left-handed doublet fermion
field, $f_R$ a right-handed singlet fermion field, and we have
suppressed flavour indices. 
The reason for focussing on this class of operators is that they are the 
only dimension-six operators that induce tree-level modifications to
Higgs--fermion couplings.
See Sec.~\ref{sec:SMEFT} for a detailed
discussion. These operators will contribute to leptonic and hadronic
dipole moments via a series of matching and renormalization-group (RG)
evolution. Our analysis takes into account the leading effects that
arise at two-loop order in the electroweak interactions, as well as
leading-logarithmic QCD corrections below the electroweak scale. We
neglect any effects of CP-odd, flavour off-diagonal operators. The
study of these effects is relegated to future work.

To understand the complex interplay between the operators we consider
requires us to vary multiple Wilson coefficients simultaneously. To
explore this computationally challenging multidimensional parameter
space, we make use of the {\small\texttt{GAMBIT}}~\cite{GAMBIT:2017yxo} global
fitting framework, to which we have added a new module that allows for
calculations of EDMs in this EFT, as well as the corresponding
experimental likelihoods. We have also expanded the existing
{\small\texttt{ColliderBit}}~\cite{GAMBIT:2017qxg} module to be able to 
determine constraints on the Wilson coefficients from measured properties 
of the Higgs boson at the LHC.

This work contains several new aspects and presents nontrivial
results. For the first time, we perform a multi-parameter fit to both
LHC and EDM data. This corresponds to a more realistic scenario than
single-parameter fits, as in a UV-complete model several CP phases are
expected to be present. Moreover, we complement and correct some of
the analytic expressions in the literature.
In addition to updating existing constraints on either individual
CP-odd Yukawa couplings (Figs.~\ref{fig:1flavour}
and~\ref{fig:1flavour_3rdgen}), or two CP-odd Yukawa couplings as in
Ref.~\cite{Chien:2015xha} (Fig.~\ref{fig:udbsscan}), we scan over up
to six CP-even / CP-odd Yukawa couplings simultaneously for the first
time (Figs.~\ref{fig:tbtauscan}~-~\ref{fig:tbscanLHC}). While this
represents still only a subset of all Yukawa operators, we argue that
with current experimental results, including more parameters will not
lead to additional significant constraints. For instance, allowing for
any CP-odd contribution to the Yukawa of a light fermion (electron, or
up and down quark) would allow to fully cancel the corresponding EDM
constraint, thus leaving only the LHC bounds in a trivial
way. However, we find several nontrivial effects. Focusing on the
heavier fermions, we find an intricate interplay between LHC and
precision constraints that weakens the EDM bounds without lifting them
fully. Not even by allowing for six independent parameters can we
cancel all EDM constraints among the third-generation fermions, and
nontrivial EDM bounds remain. A detailed discussion of these issues is
found in Sec.~\ref{sec:global}.

In our analysis, we do not take into account perturbative and hadronic
uncertainties. The effects of hadronic uncertainties on EDM bounds
have been studied in detail in Ref.~\cite{Chien:2015xha} (see also
Ref.~\cite{Jung:2013hka}), and will not be repeated here. They are
relevant mainly for bounds on CP-violating Higgs couplings to the
light quarks that arise from hadronic EDMs, as collider bounds are
nearly absent. When allowing for the presence of several Wilson
coefficients at the same time, EDM bounds tend to get canceled and
only collider bounds remain, so hadronic uncertainties are less
relevant. Note also that bounds from arising from the electron EDM
have no hadronic uncertainty. However, there is a large, previously
overlooked {\em perturbative} uncertainty regarding the bottom and
charm couplings, as discussed in Sec.~\ref{sec:EFT:below}. This
uncertainty is analogous to the case discussed in
Ref.~\cite{Brod:2018pli}, and will be studied in detail in a
forthcoming publication~\cite{BS2022}. As not all relevant
higher-order corrections are currently known, they are neglected in
this analysis.

This paper is organised as follows. In Section~\ref{sec:SMEFT} we
specify the operators that we want to constrain; namely, those that
modify the SM Yukawa couplings at tree level. We then derive the
modified Higgs--fermion couplings in the broken electroweak phase. In
Sec.~\ref{sec:EFT:below} we discuss the effective theory valid below
the electroweak scale. The analytic contributions of the SMEFT
operators to the partonic EDM of leptons and quarks are collected in
Sec.~\ref{sec:match}. Most of these results are taken from the
literature, although a few results are presented here for the first
time. Furthermore, we correct several errors in the literature. In
Sec.~\ref{sec:constraints:EDM} we summarize the contributions of the
partonic EDMs to the EDM of the physical systems that are actually
constrained by experiment. Sec.~\ref{sec:collider} describes the
collider constraints on the SMEFT operators, and contains a short
discussion of the expected contribution of operator with mass
dimension eight. Our main results are contained in
Sec.~\ref{sec:global}. We conclude in Sec.~\ref{sec:conclusions}. 
App.~\ref{sec:full:lagrangian} contains the full generic $R_\xi$-gauge Lagrangian 
in the broken phase.
In App.~\ref{sec:nir:basis} we discuss the alternative flavour basis in
the electroweak broken phase that has been used in
Ref.~\cite{Fuchs:2020uoc}. 

\section{Effective theory above the electroweak scale -- SMEFT\label{sec:SMEFT}}

In this work we consider the SM augmented with SMEFT operators that induce 
tree-level modifications to the Yukawa couplings. 
In the unbroken phase, the relevant part of the Lagrangian reads
\begin{equation}
  \begin{split}
  \Lag_{\text{Yukawa}} = 
  &- \bar{Q}_L \tilde{H} Y_u u_R + \frac{1}{\Lambda^2} (H^\dagger H)   \bar{Q}_L \tilde{H} C_{uH}' u_R\\ 
  &- \bar{Q}_L H         Y_d d_R + \frac{1}{\Lambda^2} (H^\dagger H)   \bar{Q}_L        H  C_{dH}' d_R\\
  &- \bar{L}_L H       Y_\ell \ell_R + \frac{1}{\Lambda^2} (H^\dagger H)   \bar{L}_L        H  C_{\ell H}' \ell_R
  +\text{h.c.}\,.
  \end{split}
  \label{eq:Lwarsaw}
\end{equation}
Here, $u_R$, $d_R$, $\ell_R$ denote the triplets (in generation space)
of right-handed up-, down-, and charged-lepton fields, while $Q_L$ and
$L_L$ are the corresponding left-handed triplets. In accordance, the
Yukawa matrices $Y_u$, $Y_d$, $Y_\ell$ and Wilson coefficients
$C_{uH}'$, $C_{dH}'$, $C_{\ell H}'$, are generic complex $3\times 3$
matrices.  The primes indicate that they are not necessarily couplings
in the mass-eigenstate basis. Phases in the Yukawa couplings and
Wilson coefficients can potentially induce beyond-the-SM (BSM) 
CP-violating contributions to Higgs--fermion couplings. As is well-known,
not all these complex parameters are physical, due to the freedom of
choosing the phases of the left- and right-handed fermion fields. To
isolate the physical BSM parameters, we express as many parameters as
possible in terms of observed quantities, e.g., the fermion masses and
CKM matrix elements. To this end, we rewrite the Lagrangian in the
broken phase using the linear decomposition of the Higgs field as
\begin{equation}
H =
\begin{pmatrix}
G^+\\
\tfrac{1}{\sqrt{2}} \big( v + h + iG^0\big)
\end{pmatrix} \,.
\end{equation}
The terms that induce the fermion masses are
\begin{equation}\label{eq:fermion:mass:term}
  \Lag_{\text{mass}} = - \sum_{f=u,d,\ell}
  \frac{v}{\sqrt{2}} \bar{f}_L \left(Y_f - \frac{v^2}{2\Lambda^2} C_{fH}' \right)f_R
  +\text{h.c.}\,.
\end{equation}
In analogy to the SM, we diagonalise the complete matrix in
parentheses by biunitary transformations, parameterised by the field
redefinitions
\begin{equation}\label{eq:UWrotation}
  f_L \to U_f f_L\,, \qquad f_R \to W_f f_R\,,
\end{equation}
with $f = u, d, \ell$, and $U_f$, $W_f$ complex $3 \times 3$ matrices
chosen such that after this rotation, the mass Lagrangian is
\begin{equation}
  \Lag_{\text{mass}} = - \sum_{f=u,d,\ell}
  \frac{v}{\sqrt{2}} \bar{f}_L y_f^{\text{SM}} f_R
  +\text{h.c.}\,.
\end{equation}
Here, the $y_f^{\text{SM}}$ are real and diagonal matrices with
entries that correspond to the observed fermion masses, i.e., $m_f =
\frac{v}{\sqrt{2}} y_f^{\text{SM}}$. In other words, the conditions on
$U_f$, $W_f$ read
\begin{align}
  y^{\text{SM}}_f \equiv U_f^\dagger \left(Y_f - \frac{v^2}{2\Lambda^2} C_{fH}' \right) W_f \,.
  \label{eq:ysm}
\end{align}
The kinetic terms of the fermions are also affected by the
transformation in Eq.~\eqref{eq:UWrotation}, which leads to the
appearance of the CKM matrix $V_{\text{CKM}} \equiv U_u^\dagger W_d$
in the charged gauge interactions of the left-handed quarks, in
complete analogy to the SM.

The interaction terms of the fermions with a single Higgs field from
Eq.~\eqref{eq:Lwarsaw} {\itshape after} performing the transformation
in Eq.~\eqref{eq:UWrotation} are given by
\begin{equation}\label{eq:L:h:rot}
  \begin{split}
  \Lag_{h} & = - \sum_{f=u,d,\ell}
  \frac{h}{\sqrt{2}} \bar{f}_L U_f^\dagger \left(Y_f - \frac{3v^2}{2\Lambda^2} C_{fH}' \right) W_f f_R
  +\text{h.c.} \\
  & =
  - \sum_{f=u,d,\ell}
  \frac{h}{\sqrt{2}} \bar{f}_L \left(y_f^{\text{SM}} - \frac{v^2}{\Lambda^2} C_{fH} \right)f_R
  +\text{h.c.} \,,
  \end{split}
\end{equation}
where we defined $C_{fH} \equiv U_f^\dagger C_{fH}' W_f$. We see that, in the
most general case, we can parametrise any deviation with respect to
the SM by the rotated Wilson coefficients $C_{fH}$. For later
convenience, we split these coefficients into their real and imaginary
parts,
\begin{equation}\label{eq:def:WC:rot}
  C_{fH,ij} = \text{Re}[C_{fH,ij}]+ i \, \text{Im}[C_{fH,ij}] \equiv C_{fH+,ij} + i C_{fH-,ij} \,.
\end{equation}
with $C_{fH\pm,ij}$ real. This implies that the operator combinations
proportional to the diagonal terms $C_{fH\pm,ii}$ are self-adjoint.

In the current work we consider the effect of the flavour-diagonal
contributions, $C_{fH,ii}$, for the case that they contain additional
sources of CP violation. This implies, as we shall see, that
$C_{fH-,ii}$ is non-zero. Note, however, that in the absence of any
type of flavour alignment with the SM Yukawas these operators also
induce flavour violation beyond the SM, which we do not consider in
this work. In a concrete model, we thus expect additional to
constraints from such flavour off-diagonal contributions of $C_{fH}$
(see Ref.~\cite{Alonso-Gonzalez:2021jsa} for a recent discussion).
It is, however, possible to obtain both CP-even and CP-odd
modifications in the flavour-diagonal Yukawas without any
flavour-violation beyond the SM. This corresponds to a UV scenario in
which the unitary transformations in Eqs.~\eqref{eq:ysm}
and~\eqref{eq:UWrotation} that rotate to the mass eigenstates
simultaneously diagonalise the coefficients $C'_{fH}$. In this setup
$C_{fH}$ is diagonal but not necessarily real, meaning that CP
violation beyond the SM is still possible.

In unitarity gauge, all BSM effects in our setup are contained in the 
Yukawa Lagrangian in the mass-eigenstate basis
\begin{equation}\label{eq:Lag:unit:rot}
  \begin{split}
    \Lag_{\text{Yukawa}} =\sum_f\biggl[
      \left( -m_f -m_f \frac{h}{v} \right.
      &+ \left.\frac{v^3}{2\sqrt{2}\Lambda^2} C_{fH+}
         \left(2 \frac{h}{v} +  3\frac{h^2}{v^2} + \frac{h^3}{v^3}\right) \right) \bar{f}f\\
      &+ \frac{v^3}{2\sqrt{2}\Lambda^2} C_{fH-}
         \left( 2 \frac{h}{v} +  3\frac{h^2}{v^2} + \frac{h^3}{v^3} \right) \bar{f}i \gamma_5 f\biggr]
  \end{split}
\end{equation}
Here, the sum runs over all charged fermion fields,
$f=u,d,s,c,b,t,e,\mu,\tau$, and we have traded the generation indices
on the Wilson coefficients in favour of the flavour label, $f$, since
we focus on the flavour-diagonal case. We will use this more compact
notation in the remainder of this work.  In
App.~\ref{sec:full:lagrangian} we present the corresponding Lagrangian
for generic $R_\xi$ gauge. A different basis has been used in
Ref.~\cite{Fuchs:2020uoc} to present the bounds, see
App.~\ref{sec:nir:basis}.

Often, the $\kappa$ framework is employed to parametrise
flavour-diagonal CP-violation in the Yukawas~\cite{Brod:2013cka,
  Brod:2018pli, Brod:2018lbf}.  The corresponding Lagrangian reads
\begin{equation}\label{eq:Lag:kappa}
  \Lag_{\text{Yukawa},\kappa} = - \sum_f
      \frac{y_f^{\text{SM}}}{\sqrt{2}} \kappa_f 
    \bar f\left( \cos\phi_f +  i \gamma_5 \sin\phi_f\right) f \,,
\end{equation}
with $\kappa_f$ a real parameter controlling the absolute value of the
modification and $\phi_f\in[0,2\pi)$ a CP-violating phase, such that
  $\kappa_f^{\text{SM}}=1$ and $\phi_f^{\text{SM}} = 0$ reproduces the
  SM.  Eq.~\eqref{eq:Lag:kappa} should be thought of as the
  dimension-four part of the corresponding HEFT Lagrangian in
  unitarity gauge.  As long dimension-six SMEFT is a good
  approximation, and the $h^2$ and $h^3$ interactions in
  Eq.~\eqref{eq:Lag:unit:rot} do not affect the observables
  considered, the bounds on $C_{fH\pm}$ directly translate to bounds
  on the $\kappa$-framework parameters via
\begin{align}\label{eq:kappa}
  &\kappa_f\cos\phi_f \circeq 1 - \frac{v}{\sqrt{2}m_f} \frac{v^2}{\Lambda^2} C_{fH+}\,,& 
  &\kappa_f\sin\phi_f \circeq   - \frac{v}{\sqrt{2}m_f} \frac{v^2}{\Lambda^2} C_{fH-}\,.& 
\end{align}
This is actually the case for the observables and the precision that
we consider. A possible exception is the top contribution to the LHC
observables, see the discussion in Sec.~\ref{sec:collider}.

Using the Lagrangian in Eq.~\eqref{eq:Lag:unit:rot} we will be able to
set constraints on the $C_{fH\pm}$ couplings from LHC measurements.
Low-energy probes of CP-violation, however, also place significant
constraints on these couplings. Our aim is capture the numerically
leading effects. To this end we work with a tower of effective
theories.  Apart from a few exceptions that we discuss below, the RG
evolution from $\Lambda$ to $\muew$ within SMFET can be neglected as
(most) Yukawa operators do not mix into the CP-violating dipoles for
the electron and light-quarks, relevant for EDMs. In this case the
leading effects are obtained at the electroweak scale by matching to
the effective theory in which the heavy degrees of the SM are
integrated out, and by subsequently running to the hadronic scale.  In
contrast, the few contributions that are induced from (two-loop)
mixing within SMEFT will come with a UV-sensitive logarithm, $\log
M_h/\Lambda$.

\section{Effective theory below the electroweak scale\label{sec:EFT:below}}

We will extract constraints on the coefficients in
Eq.~\eqref{eq:Lag:unit:rot} from the experimental bounds on the
electron, neutron, and mercury EDMs. We will make the assumption that
these EDMs are induced solely by the electron and partonic
CP-violating electric and chromoelectric dipole operators, and the
purely gluonic CP-violating Weinberg operator \cite{Weinberg:1989dx}
with coefficients $d_e$, $d_q$, $\tilde d_q$, and $w$, respectively.
Therefore, we neglect (subleading) contributions from the matrix
elements of four-fermion operators.  Here, $q$ collectively denotes
light quarks, $u$, $d$, $s$. The charm-quark is typically not included
since presently there are no lattice computations for the hadronic
matrix elements of its operators.  The above coefficients are
traditionally defined via the effective, CP-odd Lagrangian valid at
hadronic energies $\muhad\simeq 2\,$GeV~\cite{Engel:2013lsa},
\begin{multline} \label{eq:LeffN}
  \Lag_{\rm eff} = 
  - d_e  \frac{i}{2}  \bar e \sigma^{\mu\nu} \gamma_5 e  F_{\mu\nu} 
  - d_q  \frac{i}{2}  \bar q \sigma^{\mu\nu} \gamma_5 q  F_{\mu\nu} 
  - \tilde d_q  \frac{ig_s}{2}  \bar q \sigma^{\mu\nu} T^a \gamma_5 q  G_{\mu\nu}^a 
  + \frac{1}{3} w f^{abc}  G_{\mu \sigma}^a G_{\nu}^{b, \sigma} \widetilde G^{c, \mu \nu}\,.
\end{multline}
Here, $\sigma^{\mu\nu} = \tfrac{i}{2} [\gamma^\mu, \gamma^\nu]$, $T^a$
are the generators of SU$(n_c)$ in the fundamental representation
normalised as Tr$[T^a,T^b]= \delta^{ab}/2$, $n_c=3$ is the number of
colors, and we collect our conventions for the field strength and its
dual below.  The contributions from the Weinberg operator turn out to
be subdominant because of its small nuclear matrix
elements~\cite{Pospelov:2005pr, Engel:2013lsa}, but we keep them for
completeness.

The partonic dipole moments $d_q$ and $\tilde{d}_q$ as well as the
coefficient $w$ are obtained from the SMEFT Wilson coefficients in
Eq.~\eqref{eq:Lag:unit:rot} via a sequence of matching at the weak
scale, and the bottom- and charm-flavour thresholds, as well as the RG
evolution between each scale, as outlined in
Refs.~\cite{Degrassi:2005zd, Hisano:2012cc, Brod:2013cka,
  Brod:2018pli, Panico:2018hal}. The resummation of logarithms is
phenomenologically relevant for the hadronic dipole operators that mix
under QCD. In the current analysis, we work at the leading-logarithmic
order for quark operators. QCD corrections are also large for the
bottom- and charm-quark contributions to the electron EDM. They are
currently unknown~\cite{BS2022} and are therefore not included in our
analysis.

To perform the sequence of matching and RG evolution connecting the
electroweak-scale Lagrangian, Eq.~\eqref{eq:Lag:unit:rot}, with the
hadronic-scale one, Eq.~\eqref{eq:LeffN}, requires the full
flavour-conserving, CP-odd effective Lagrangian below the electroweak
scale up to mass dimension six. In the conventions and basis of
Ref.~\cite{Brod:2018pli}, adapted from Ref.~\cite{Hisano:2012cc}, it
reads:
\begin{equation}\label{eq:Leff}
\begin{split}
\Lag_\text{eff} =
  - \sqrt{2} G_F\, \bigg(
	& \sum_{q\neq q'}
	  \bigg[ \sum_{i=1,2} C_i^{qq'}  O_i^{qq'} + \frac{1}{2} \sum_{i=3,4} C_i^{qq'}  O_i^{qq'} \bigg] \\
        & + \sum_q  \!\!\sum_{i=1,\ldots,4}\!\! C_i^q   O_i^q  + C_w  O_w 
          + \sum_\ell C^\ell_3 O_3^\ell \bigg ) \,,
\end{split}
\end{equation}
where the sums run over all active quarks with masses below the weak
scale, e.g., $q,q' = u,d,s,c,b$ in the five-flavour theory.  The
linearly independent operators are\footnote{The definition of the
  $O_3^{qq'}$, $O_4^{qq'}$ operators in terms of the $\epsilon$-tensor
  is convenient when going beyond leading-logarithmic approximation,
  as they remain selfadjoint also in $d=4-2\epsilon$ dimensions. It is
  thus convenient to also define $O_2^q$ and the dipoles $O_3^q$ and
  $O_4^q$ in an analogous manner. To express the operators in the more
  conventional $i\sigma^{\mu\nu}\gamma^5$ notation one can use the
  $d=4$ relation
\begin{equation*}
  i\sigma^{\mu\nu}\gamma_5 \overset{[d=4]}{=} \frac{1}{2}\epsilon^{\mu\nu\rho\sigma}\sigma_{\rho\sigma}\,,
\end{equation*}
where $\epsilon_{0123}=-\epsilon^{0123}=1$. For details concerning the
next-to-leading-logarithmic analysis see Ref.~\cite{Brod:2018pli}.}
\begin{align}
\label{eq:op:qq':1:2}
O_1^{qq'} & = (\bar q q) \, (\bar q' \, i \gamma_5 q')\,,&
O_2^{qq'} & = (\bar q\, T^a  q) \, (\bar q' \, i \gamma_5 T^a q') \,,\\[0.5em]
\label{eq:op:qq':3:4}
O_3^{qq'} & = \frac{1}{2} \epsilon^{\mu\nu\rho\sigma} (\bar q \sigma_{\mu\nu} q)     \, (\bar q' \sigma_{\rho\sigma} q')\,,&
O_4^{qq'} & = \frac{1}{2} \epsilon^{\mu\nu\rho\sigma} (\bar q \sigma_{\mu\nu} T^a q) \, (\bar q' \sigma_{\rho\sigma} T^a q')\,,\\[0.5em]
\label{eq:op:q}
O_1^q & = (\bar q q) \, (\bar q \, i \gamma_5 q) \,,&
O_2^q &= \frac{1}{2}\epsilon^{\mu\nu\rho\sigma}(\bar q \sigma_{\mu\nu} q) \, (\bar q \, \sigma_{\rho\sigma} q) \,,\\[0.5em]
\label{eq:dipoles}
O_3^q & = \frac{eQ_q}{2} \frac{m_q}{g_s^2} \, \bar q \sigma^{\mu \nu} q \, \widetilde{F}_{\mu \nu} \,,&
O_4^q & = -\frac{1}{2} \, \frac{m_q}{g_s} \, \bar q \sigma^{\mu \nu} T^a q \, \widetilde{G}^a_{\mu \nu} \,,\\[0.5em]
O_3^\ell &= \frac{eQ_\ell}{2} \frac{m_\ell}{e^2} \, \bar \ell \sigma^{\mu \nu} \,i \gamma_5 \ell \, F_{\mu \nu} \,,&
O_w     & = -\frac{1}{3\,g_s} f^{abc} \, G_{\mu \sigma}^a G_{\nu}^{b, \sigma} \widetilde G^{c, \mu \nu} \,.\label{eq:op:weinberg}
\end{align}
The powers of $e$ and $g_s$ in the operator definitions above are
fixed such that all Wilson coefficients have the same $\hbar$
dimension, thus making the calculation of operator mixing more
transparent.

$C_w$ as well as some of the anomalous dimensions which control
operator mixing depend on the conventions for the covariant
derivatives for quarks and leptons, the field strength tensor, and the
dual tensors. Our conventions (same as in Ref.~\cite{Brod:2018pli})
read
\begin{align}
  D^q_\mu    &\equiv \partial_\mu - i g_sT^a G^a_\mu + i e Q_q    A_\mu\,, &
  D^\ell_\mu &\equiv \partial_\mu                    + i e Q_\ell A_\mu\,,\\
  G_{\mu\nu}^a&\equiv \partial_\mu G_\nu^a - \partial_\nu G_\mu^a + g_s f^{abc} G_\mu^b G_\nu^c,&
  \widetilde{G}^{a,\mu\nu}&\equiv\frac{1}{2}\epsilon^{\mu\nu\rho\sigma}G^a_{\rho\sigma}\,,\qquad
  \widetilde{F}^{\mu\nu}\equiv\frac{1}{2}\epsilon^{\mu\nu\rho\sigma}F_{\rho\sigma}\,,\nonumber
\end{align}
with {\itshape (not unimportantly)} $\epsilon_{0123}=-\epsilon^{0123}=1$.

The comparison of Eqs.~\eqref{eq:LeffN} and \eqref{eq:Leff} gives the
relation between $d_e$, $d_q$, $\tilde{d}_q$, $w$ and the Wilson
coefficients of the three-flavour EFT evaluated at the hadronic scale
$\muhad = 2\,$GeV:
\begin{equation}
\label{eq:ddw}
\begin{split}
  d_e              & = \phantom{+}\sqrt{2} G_F \,\frac{eQ_e}{4\pi \alpha} m_e \,C_3^e \,,\\
d_q (\muhad)       & = \phantom{+}\sqrt{2} G_F \,\frac{eQ_q}{4 \pi \alpha_s}      m_q \, C_3^q (\muhad) \,,\\
\tilde d_q(\muhad) & =          - \sqrt{2} G_F \,\frac{1}   {4 \pi \alpha_s}      m_q \, C_4^q (\muhad) \,,\\
w (\muhad)         & = \phantom{+}\sqrt{2} G_F \,\frac{1}{g_s} \, C_w (\muhad) \,.
\end{split}
\end{equation}
Note that $C_3^e$ is scale independent with regard to the strong
interaction, i.e., $C_3^e(\muhad)\simeq C_3^e(\muew)$ with
$\muew\simeq 100$\,GeV.  The procedure to calculate $C_3^e$,
$C_3^q(\muhad)$, $C_4^q(\muhad)$, and $C_w (\muhad)$ depends on the
flavour quantum numbers of the SMEFT operators.  In the following
section, we describe the different cases and present the corresponding
results.

\section{Electroweak matching and RG evolution to the hadronic scale\label{sec:match}}

In this section we discuss how to obtain the Wilson coefficients of
the EFT below the electroweak scale as a function of the Yukawa SMEFT
Wilson coefficients. Since all operators in Eq.~\eqref{eq:Leff} are CP
odd, all contributions are proportional to the CP-odd couplings
$C_{fH-}$.  The final results for the required initial conditions
$C_{1}^{q\qp}(\muew)$, $C_{1}^{q}(\muew)$, $C_w(\muew)$,
$C^e_3(\muew)$, $C^q_3(\muew)$, and $C^q_4(\muew)$ are collected below
in Eqs.~\eqref{eq:C1tree}, \eqref{eq:Cwew}, \eqref{eq:C3eew},
\eqref{eq:C3qew}, and~\eqref{eq:C4qew}, respectively.  These are all
the initial conditions required for the leading-log QCD analysis.

\bigskip

We begin the discussion with contributions that are UV sensitive,
i.e., proportional to $\log(\muew/\Lambda)$.  There are two equivalent
ways of obtaining them.  One way is to identify those SMEFT Yukawa
operators $O_{fH}$ that mix into the SMEFT dipole operators $O_{fW}$,
$O_{fB}$ and $O_{qG}$. The mixing induces dipole Wilson coefficient
proportional to the leading UV logarithm $\log\muew/\Lambda$, and the
subsequent tree-level matching onto the Lagrangian Eq.~\eqref{eq:Leff}
potentially induces non-zero coefficients $C^e_3(\muew)$,
$C^q_3(\muew)$, and $C^q_4(\muew)$. In fact, only the following cases
occur: the operator $O_{eH}$ mixes at the two-loop, electroweak level
into the leptonic dipole operators
\begin{align}
O_{eW} &= \bar{L}_L  \sigma^a \sigma^{\mu\nu}e_R H W^a_{\mu\nu}\,, &
O_{eB} &= \bar{L}_L  \sigma^{\mu\nu}e_R H B_{\mu\nu}\,.
\end{align}
Analogously, the Yukawa operators $O_{qH}$ with $q=u,d,s$ mix into the
quark dipole operators
\begin{align}
O_{qW} &= \bar{Q}_L\sigma^a \sigma^{\mu\nu}q_R \hat{H} W^a_{\mu\nu}\,,&
O_{qB} &= \bar{Q}_L \sigma^{\mu\nu}q_R         \hat{H} B_{\mu\nu}\,, 
\end{align}
with $\hat{H}\equiv H$ for $q=d,s$ and $\hat{H}\equiv \tilde{H}$ for
$q=u$. Here, $\sigma^a$ are the Pauli matrices acting in $SU(2)_L$
space. (Here, we considered only mixing into dipole operators with
light fermions.) The two-loop mixing has been calculated for the
electron case in Ref.~\cite{Panico:2018hal}.

\begin{figure}[t]
  \begin{center}
    \includegraphics[]{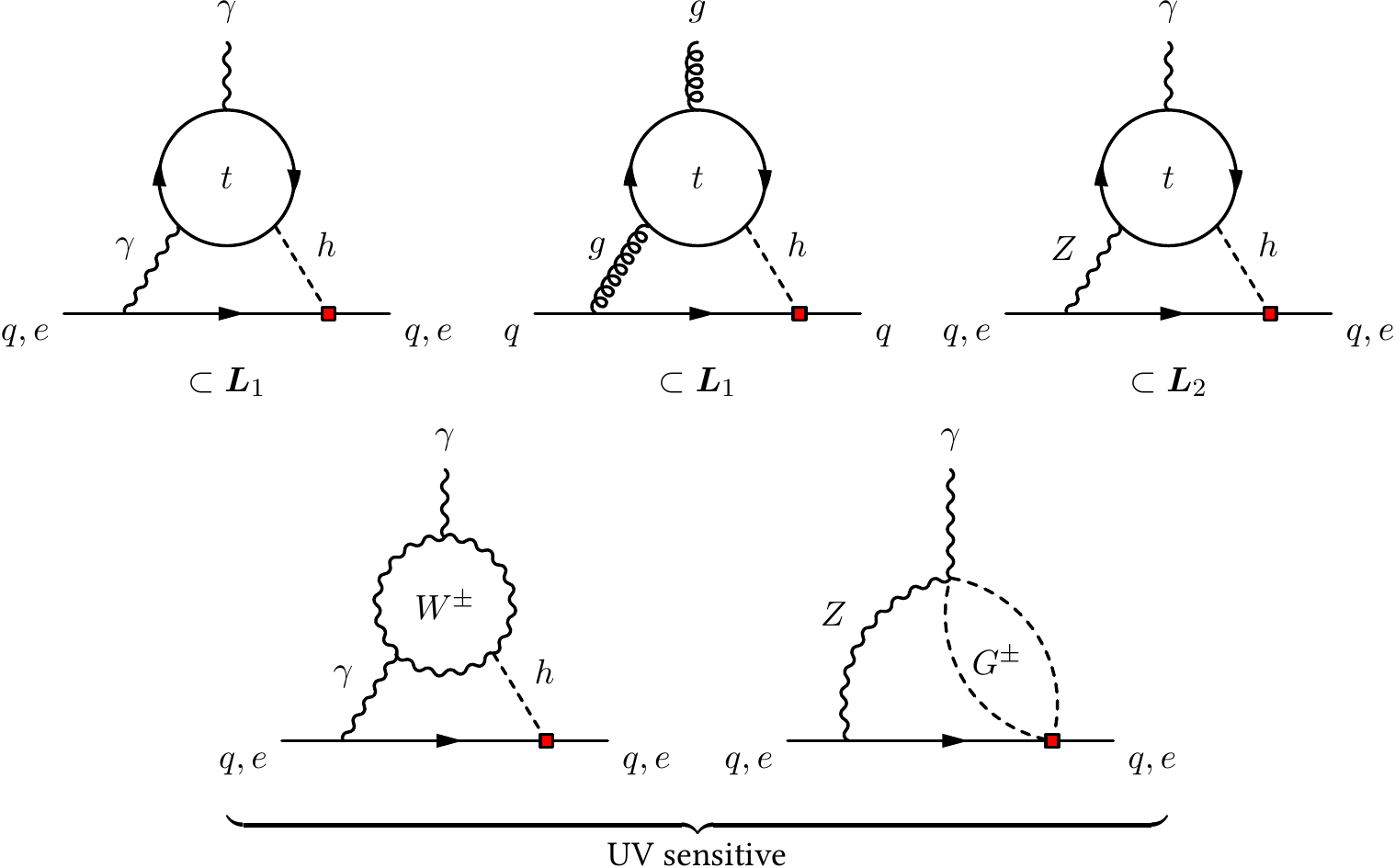}
    \caption{Representative Barr--Zee-type Feynman diagrams
      contributing to the two-loop matching of dipole operators at the
      electroweak scale.  The label under the diagrams indicates the
      loop function (${\boldsymbol L}_1$ and ${\boldsymbol L}_2$) to
      which the corresponding diagram contributes.  The red square
      vertex indicates a CP-violating Higgs Yukawa, $C_{fH-}$.
      Electroweak diagrams like those in the second row are
      responsible for the UV sensitivity of contributions proportional
      to $C_{eH-}$ and $C_{qH-}$ for the electron and light-quark EDM
      operators, respectively.  Diagrams with dimension-five couplings
      (lower right) are required to obtain a gauge-invariant result.
      See main text for details.
      \label{fig:feynmanL1L2UV}}
  \end{center}
\end{figure}

An alternative way of obtaining these $\log\muew/\Lambda$ terms is to
directly perform the two-loop electroweak matching calculation to
extract $C_3^f(\muew)$, for the light-fermion flavours $f=e,u,d,s$
(see representative diagrams in the lower panels in
Fig.~\ref{fig:feynmanL1L2UV}).  For these coefficients, the
contributions to the matching proportional to the SMEFT coefficients
$C_{fH-}$ are divergent and thus include a term proportional to the
UV-sensitive logarithm $\log M_h/\Lambda$. We performed the matching
calculation explicitly and find that only the contributions to
$C_3^f(\muew)$ are UV sensitive, while the matching for $C_4^q(\muew)$
is UV finite. To obtain gauge-independent results in this calculation,
it was necessary~\cite{Altmannshofer:2020shb} to include the
dimension-five vertices in the Lagrangian in the broken phase (see
App.~\ref{sec:full:lagrangian}) that were missed in
Ref.~\cite{Altmannshofer:2015qra}.  The UV-divergent contributions are
a direct consequence of including these dimension-five vertices.
The corresponding logarithmic terms in our results for $C_3^e(\muew)$
are in agreement with the anomalous dimension presented in
Ref.~\cite{Panico:2018hal}. The UV-sensitive logarithms for the quark
case are presented here for the first time.

When electroweak UV logarithms are induced, the electroweak
non-logarithmic terms in the matching at $\muew$ that accompany them
are scheme dependent and formally part of the
next-to-leading-logarithmic (NLL) approximation.  They should not be
included in the analysis and we thus disregard them.

\bigskip

For those contributions that are not UV sensitive, we directly perform
the matching of the SMEFT Lagrangian Eq.~\eqref{eq:Lag:unit:rot} onto
the effective Lagrangian Eq.~\eqref{eq:Leff}, integrating out the
heavy degrees of the SM (top quark, Higgs, and the $W$ and $Z$
bosons).  In all calculations we employ a general $R_\xi$ gauge for
all gauge bosons and have verified the independence of our results of
the gauge-fixing parameters.  All calculations were performed using
self-written {\small\texttt{FORM}}~\cite{Vermaseren:2000nd} routines,
implementing when necessary the two-loop recursion presented in
Refs.~\cite{Davydychev:1992mt, Bobeth:1999mk}. The amplitudes were
generated using {\small\texttt{QGRAF}}~\cite{Nogueira:1991ex} and
{\small\texttt{FeynArts}}~\cite{Hahn:2000kx}.

\begin{figure}[t]
  \begin{center}
    \includegraphics[]{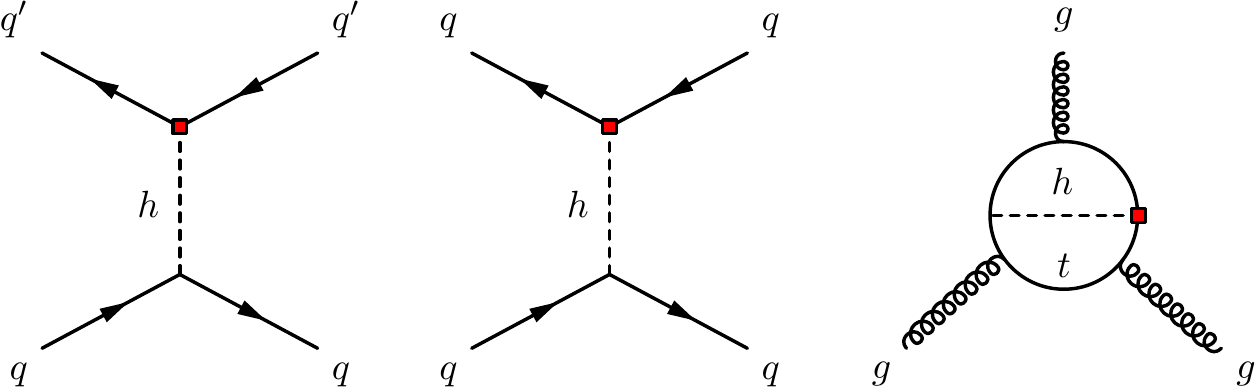}
    \caption{Sample Feynman diagrams contributing to the tree-level matching
      of four-quark operators (first two diagrams) and the two-loop matching
      inducing the Weinberg operator (third diagram).
      The red square vertex indicates a CP-violating Higgs Yukawa, $C_{fH-}$.
    \label{fig:feynmantreewein}}
  \end{center}
\end{figure}

At tree level and leading order in the $1/\Lambda$ expansion,
integrating out the Higgs induces the following non-zero initial
conditions for four-quark operators (see first two diagrams in Fig.~\ref{fig:feynmantreewein})
\begin{equation}\label{eq:C1tree}
  C_{1}^{qq'}(\muew) = \frac{m_{q}m_{q'}}{M_h^2} \frac{v}{\sqrt{2}m_{q'}} \frac{v^2}{\Lambda^2} C_{q'H-} 
  \,, \qquad
  C_{1}^{q}(\muew)   = \frac{m_{q}^2}{M_h^2} \frac{v}{\sqrt{2}m_{q}} \frac{v^2}{\Lambda^2} C_{qH-} \,,
\end{equation}
with $q,q'=u,d,s,c,b$.  In principle, also the analogous operators
with leptons are induced ($\bar q q\bar\ell\ell$,
$\bar\ell'\ell'\bar\ell\ell$).  However, since we will rely on a
fixed-order computation to predict $d_e\propto C_3^e$ (see below),
these operators do not enter our analysis.

The matching at the electroweak scale also induces the Weinberg and
dipole operators, but only at the two-loop level.  The Weinberg
operator obtains a two-loop initial condition at $\muew$ only from the
top-quark operator (see third diagram in Fig.~\ref{fig:feynmantreewein}):
\begin{equation}
  \label{eq:Cwew}
\begin{split}
  C_w(\muew) & = \frac{\alpha_s^2}{(4 \pi)^2} \frac{v}{\sqrt{2}m_t} \frac{v^2}{\Lambda^2} C_{tH-}
\frac{\xth}{2(1 - 4 \xth)^3} \bigg\{ 3 - 14 \xth + 8 \xth^2  \\ & \qquad \qquad
  + 6 \xth (1 - 2 \xth + 2 \xth^2) \Phi\left(\frac{1}{4\xth}\right)
+ \big(2 + 10 \xth - 12 \xth^2 \big) \log \xth \bigg\}\,,
\end{split}
\end{equation}
with $\Phi(z)$ defined in Eq.~\eqref{eq:Phiz}.  This result agrees
with Ref.~\cite{Brod:2013cka}, but is, as far as we are aware,
presented here for the first time in a non-parametric form.

\begin{figure}[t]
  \begin{center}
    \includegraphics[]{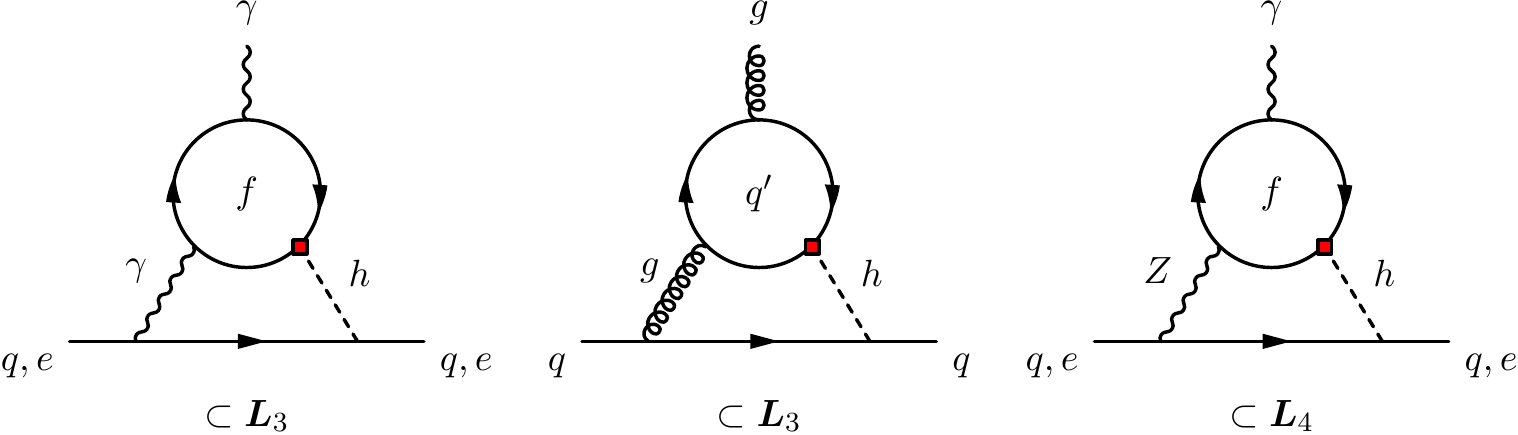}\\[2em]
    \includegraphics[]{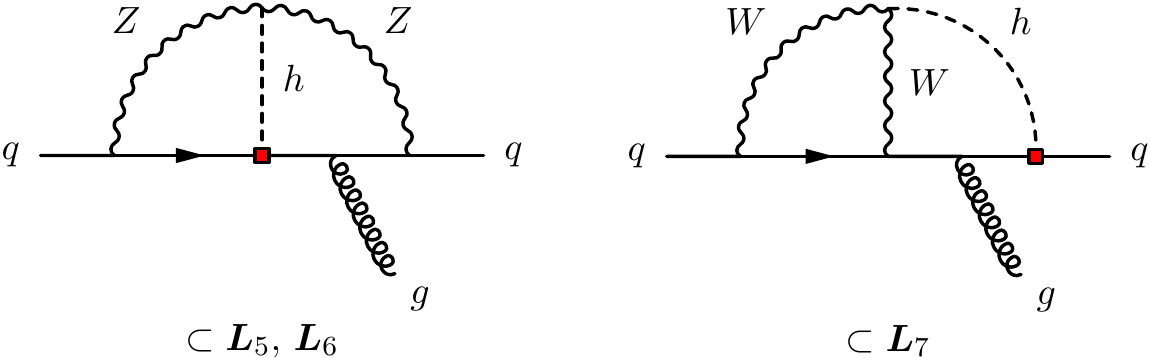}
    \caption{Representative Barr--Zee-type Feynman diagrams contributing to the two-loop matching
      of dipole operators at the electroweak scale. 
      The label under the diagrams indicates the loop function 
      (${\boldsymbol L}_3$ -- ${\boldsymbol L}_7$) to which the corresponding diagram 
      contributes.
      The red square vertex indicates a CP-violating Higgs Yukawa, $C_{fH-}$.
      Here, $f$ denotes fermions and $q'$ quarks.
      The electroweak contributions to the chromo EDM operator from 
      diagrams like the ones in the second row are UV finite as opposed to the ones
      in Fig.~\ref{fig:feynmanL1L2UV}.
      See main text for details.
    \label{fig:feynmanL3L4L5L6L7}}
  \end{center}
\end{figure}
The electron photon dipole ($O^e_3$), and light-quark photon and gluon
dipoles ($O^q_3$, $O^q_4$ with $q=u,d,s$) all receive two-loop initial
conditions by integrating out the top quark and the Higgs, $W$, and
$Z$ bosons:
\begin{align}
  C_3^e(\muew) &= \frac{\alpha^2}{(4\pi)^2} \sum_{f'=e,\mu,\tau,b,c,t}   \frac{v}{\sqrt{2}m_{f'}}\,\frac{v^2}{\Lambda^2}\,C_{f'H-}\,A^e[f']\,,\label{eq:C3eew}\\
  C_3^q(\muew) &= \frac{\alpha\alpha_s}{(4\pi)^2} \sum_{f'=q,\tau,t} \frac{v}{\sqrt{2}m_{f'}}\,\frac{v^2}{\Lambda^2}\,C_{f'H-}\,A^q[f']\,,\label{eq:C3qew}\\
  C_4^q(\muew) &= \left( \sum_{q'=q,t}
    \frac{\alpha_s^2}    {(4\pi)^2}\,B_s^q[q']
  + \frac{\alpha\alpha_s}{(4\pi)^2}\,B_{\text{ew}}^q[q]\right)\,\frac{v}{\sqrt{2}m_{q'}}\,\frac{v^2}{\Lambda^2}\,C_{q'H-}\,,\label{eq:C4qew}
\end{align}
with the precise decomposition of the coefficient functions $A^f$, $B_s^q$, 
and $B_\text{ew}^q$ discussed below.

A few clarifications regarding Eqs.~\eqref{eq:C3eew}--\eqref{eq:C4qew}
are in order. We calculate the electroweak matching in fixed-order
perturbation theory whenever there are no QCD corrections that are
enhanced by large logarithms related to light fermion masses. This is
the case for all diagrams with top-quark loops, and the diagrams with
lepton loops contributing to the electron dipole, i.e. the
contributions $C_3^e(\muew)$ that are proportional to $C_{\ell H-}$.
(We also neglect the running of the electromagnetic coupling
constant.)  Therefore, the coefficients $C_{3/4}^q(\muew)$ do not
receive threshold contributions from virtual quarks other than the top
quark, since those are properly included via the RG-mixing of the
four-quark operators (Eq.~\eqref{eq:C1tree}) into $C_{3/4}^q$, see
below. In principle, also the contributions to $C_3^e(\muew)$ of all
quarks other than the top should be obtained from the RG evolution;
however, the corresponding mixed QED-QCD RG evolution is currently
unknown\footnote{The QCD corrections to the bottom and charm
  contributions are naively estimated to be large (factor of a few in
  $C_3^e$) but require a more complex calculation~\cite{BS2022}.}.
Thus, we temporarily include the contribution of bottom and charm
quarks to $C_3^e(\muew)$ with a fixed-order calculation, with the
understanding that corrections to these results may be large. On the
other hand, we do not include the corresponding fixed-order results
for the light-quark ($u$, $d$, $s$) contributions to
$C_3^e(\muew)$. In addition to the issue of the large unkown QCD
corrections, these contributions are suppressed by the light-quark
masses. Therefore, it is not $C_{3}^e$ but the $C_{3/4}^q$
coefficients that provide the numerically leading constraints on the
$C_{qH-}$ couplings for light quarks.
Similarly, we do not include the contributions of the electron and
muon to $C_3^q(\muew)$. The appearance of their masses (that are much
smaller than the lowest scale, $\muhad$, where these operators can be
meaningfully defined) in the logarithms would spuriously enhance the
corresponding bounds by large factors. These contributions should be
taken into account properly by using the RG. The tau, on the other
hand, has a mass large enough to justify a fixed-order calculation as
an estimate and is thus included. Note, however, that numerically the
by far strongest bounds on the coefficients of all three leptons arise
from the electron EDM, and their contributions to the hadronic EDMs do
not play a role in our analysis, given current experimental data.

The decomposition of the coefficients $A^f$, $B_s^q$, and $B_\text{ew}^q$ 
in Eqs.~\eqref{eq:C3eew}--\eqref{eq:C4qew} are
\begin{align}
  A^f[f'] &=
  \begin{cases}
    \underbrace{4 n_c Q_{t}^2                     {\boldsymbol L}_1}_{t\gamma-\text{Barr--Zee}}
    \quad+ 
    \underbrace{4 n_c Q_t Q_f^{-1} v_{Zf} v_{Zt}  {\boldsymbol L}_2}_{tZ-\text{Barr--Zee}}
    \quad- 
    \underbrace{\frac{4Q_f\mp 1}{Q_f c_w^2} \log\frac{M_h^2}{\Lambda^2}}_{\text{ew UV sensitivity}}
    &\text{for}~f= f'\\
    \underbrace{4 n_c[f']Q_{f'}^2                        {\boldsymbol L}_3}_{f'\gamma-\text{Barr--Zee}} 
    \quad+ 
    \underbrace{4 n_c[f']Q_{f'} Q_f^{-1} v_{Zf} v_{Zf'}  {\boldsymbol L}_4}_{f'Z-\text{Barr--Zee}}
    &\text{for}~f\neq f'
  \end{cases}\\
  B_s^q[q'] &=
  \begin{cases}
    \underbrace{2 {\boldsymbol L}_1}_{tg-\text{Barr--Zee}}
    &\text{for}~q= q'\\
    \underbrace{2 {\boldsymbol L}_3}_{tg-\text{Barr--Zee}}
    &\text{for}~q\neq q' = t
  \end{cases}\\[1em]
  B_{\text{ew}}^q[q] &=
   \underbrace{
   (v^2_{Zq} - a^2_{Zq}) {\boldsymbol L}_5   + 
   (v^2_{Zq} + a^2_{Zq}) {\boldsymbol L}_6}_{\text{ew-finite  $qZ$ contribution}}
   \underbrace{+\frac{1}{s_w^2}                                {\boldsymbol L}_7}_{\text{ew-finite  $qW$ contribution}}
\end{align}
with $n_c[t]=n_c[q]=n_c$ and $n_c[\ell]=1$.  $v^Z_f \equiv \frac{1}{2
  s_w c_w}(T^f_3 - 2 Q_f s_w^2)$ and $a^Z_f \equiv \frac{1}{2 s_w
  c_w}T^f_3$ correspond to the vector and axial-vector coupling of the
$Z$ boson to a fermion, respectively.
In Fig.~\ref{fig:feynmanL1L2UV} and Fig.~\ref{fig:feynmanL3L4L5L6L7} we
show representative diagrams contributing to $\boldsymbol{L}_1$--$\boldsymbol{L}_7$.
As discussed, only $A^e[e]$ and
$A^q[q]$ contain UV sensitive electroweak contributions proportional
to $\log M_h^2/\Lambda^2$, thus all other scheme-dependent finite
electroweak threshold corrections have been dropped in $A^e[e]$ and
$A^q[q]$. We have, however, kept the terms from top-quarks loops
(proportional to $\boldsymbol{L}_1$ and $\boldsymbol{L}_2$) as they
depend on the top-quark Yukawa and as such are independent of the
scheme-dependent electroweak corrections.  The loop functions
($\boldsymbol{L}_i$) read
\begin{align}
 {\boldsymbol L}_1 &=  2\xth\left( 2 + \log \xth \right) +\xth \left(1 - 2\xth\right) \Phi\left(\frac{1}{4\xth}\right)\,,\\
 %%%%%
 {\boldsymbol L}_2 &=       \frac{\xtz\xth}{\xtz-\xth} \bigg[  (1-2\xth) \Phi\left(\frac{1}{4\xth}\right) 
                             - 2 \log\xhz  - (1-2\xtz) \Phi\left(\frac{1}{4\xtz}\right)\bigg]\,,\\
 %%%%%
 {\boldsymbol L}_3 &=     \xfph\Phi\left(\frac{1}{4 \xfph} \right) 
                          \quad\overset{\xfph\ll1}{\simeq}\quad 
                          \xfph\left(\log^2 \xfph + \frac{\pi^2}{3}\right)\,, \label{eq:L3}\\[1em]
 %%%%%
 {\boldsymbol L}_4 &= 
       \frac{\xfpz\xfph}{\xfpz-\xfph}
                 \bigg[  \Phi\left(\frac{1}{4\xfpz}\right) 
                      - \Phi\left(\frac{1}{4\xfph}\right)\bigg]\\[1ex]
     & \quad \overset{\xfpz,\xfph\ll1}{\simeq}
     \frac{\xfpz\xfph}{\xfpz-\xfph}
                        \big( \log^2 \xfph - \log^2\xfpz \big)\,,\\
 %%%%%
 {\boldsymbol L}_5 &= \frac{1}{3} \bigg[ \big(3\xhz^3-18\xhz^2+24\xhz\big) \Phi\left(\frac{\xhz}{4}\right)
                  + 6 \xhz \left( 1 + 2 \log \xhz \right)\nonumber\\
  & \qquad \qquad + \big(12\xhz^3-48\xhz^2+12 \xhz\big) \text{Li}_2(1 - \xhz)\nonumber\\
  & \qquad \qquad + \big(\xhz^3-4\xhz^2\big) \pi^2
     + \big(3\xhz^3-12\xhz^2\big) \log^2 \xhz \bigg] \,,\\
%%%%%%%%%%%%%%%%%%%%%%%%%%%%%%%%%%
 {\boldsymbol L}_6 &= \frac{2}{9} \xzh^2
           \bigg[ \big(3\xhz^3-6\xhz^2-24\xhz\big) \Phi\left(\frac{\xhz}{4}\right)\nonumber\\
  & \qquad \qquad \qquad + \big(6 \xhz^2 + 24 \xhz \big) \log \xhz
                         + \big(6 \xhz^2 - 24 \xhz \big) \nonumber\\
  & \qquad \qquad \qquad + \big(6\xhz^3-18\xhz-24 \big) \text{Li}_2(1 - \xhz)
           + \big(3\xhz + 4 \big) \pi^2 \bigg] \,,\\
%%%%%%%%%%%%%%%%%%%%%%%%%%%%%%%%%%
  {\boldsymbol L}_7 &= 
  \frac{1}{18} \xhw \bigg[ \big(3-6\xwh-24\xwh^2\big) \Phi\left(\frac{\xhw}{4}\right)
                  + \big(3\xwh^2 + 4\xwh^3 \big) \pi^2 \nonumber\\
  & \qquad \qquad \qquad - \big(6 \xwh + 24\xwh^2 \big) \log \xwh
                         + \big(6 \xwh - 24\xwh^2 \big) \nonumber\\
  & \qquad \qquad \qquad + \big(12\xwh^3+9\xwh^2-3\big)
    \big(2\text{Li}_2(1 - \xwh) + \log^2 \xwh \big) \bigg] \,,
\end{align}
where we defined the mass ratios $x_{ij} \equiv M_i^2 / M_j^2$.  For
the case of small fermion masses, i.e., when $x\ll 1$, we have
expanded for convenience the $\Phi(z)$ function.  Our loop functions
$\boldsymbol{L}_5$, $\boldsymbol{L}_6$, $\boldsymbol{L}_7$ correct a
global factor of $\sqrt{2}$ with respect to the corresponding ones in
Ref.~\cite{Brod:2018lbf} and a typographical sign mistake in the
$x^3_{hZ}$ coefficient of $\text{Li}_2(1-\xhz)$ in $\boldsymbol{L}_5$
in the same reference. The loop function $\boldsymbol{L}_2$ implicitly
corrects an error in Eq.~(A.5) of Ref.~\cite{Panico:2018hal}, where
two logarithms seem to have been incorrectly added.

The function $\Phi(z)$ in the results above is given
by~\cite{Davydychev:1992mt}
\begin{equation}
  \label{eq:Phiz}
  \Phi(z) = 
  \begin{cases}
    4          \sqrt{\frac{z}{1-z}} \text{Cl}_2 \bigl(2 \arcsin(\sqrt{z})\bigr) & \text{for}~0\leq z<1\,,\\[1em]
    \frac{1}{\lambda}\left( -4 \,\text{Li}_2 (\frac{1-\lambda}{2}) + 2 \log^2(\frac{1-\lambda}{2}) - \log^2 (4z) + \frac{\pi^2}{3} \right) &\text{for}~z>1\,,
  \end{cases}
\end{equation}
where $\lambda  \equiv \sqrt{1-1/z}$, and the dilogarithm and Clausen function are defined as
\begin{align}
  &{\rm Li}_2 (x) = -\int_0^x du \, \log (1-u)/u&
  &\text{and}&
  &\text{Cl}_2 (\theta) = - \int_0^\theta dx \log |2 \sin (x/2)| \,,&
\end{align}
respectively.

\bigskip

Having computed the initial conditions at the electroweak scale, we
perform the QCD RGE evolution from $\muew$ to the hadronic scale
$\muhad = 2\,$GeV (see Refs.~\cite{Brod:2013cka, Brod:2018pli} for
details).  For operators with quarks the RGE evolution resums large
QCD logarithms and accounts for mixing among different operators.  In
the case of light-quark SMEFT operators, the tree-level induced
four-quark operators ($O_{1}^{qq'}$ and $O_{1}^q$) mix at one-loop
under QCD into the quark dipoles. Nevertheless, the contributions to
$C_3^q(\muew)$ and $C_4^q(\muew)$ of two-loop diagrams with top-quark
and $Z$-boson loops provide the numerically dominant
effect~\cite{Weinberg:1989dx} as the four-fermion operators are
additionally suppressed by an extra light-quark mass (see
Eq.~\eqref{eq:C1tree}).

The situation is different for contributions from bottom- and
charm-quark SMEFT operators.  Here the leading contribution to
partonic dipoles ($d_q$, $\tilde{d}_q$) and the Weinberg operator
($w$) are induced by mixing during the RG evolution. The main reason
is that the nuclear matrix elements of bottom and charm dipole
operators are tiny.
Therefore, diagrams like the ones in Fig.~\ref{fig:feynmanL1L2UV} do not contribute, 
i.e., bottom and charm
quarks only enter as virtual particles in loop diagrams. At the same
time, their mass is significantly below the electroweak scale. For
this reason, a tree-level matching at the electroweak scale and the
subsequent one-loop RG evolution \cite{Hisano:2012cc}, which must
include the mixing of four-fermion operators into dipole and Weinberg
operators, is sufficient (and necessary) to obtain the
leading-logarithmic result. The two-loop calculation has been
performed in Ref.~\cite{Brod:2018pli} but is not used in this work, as
perturbative uncertainties and higher-order corrections are generally
neglected in our analysis (see the discussion in
Sec.~\ref{sec:had:error}).

\section{Electric dipole moments\label{sec:constraints:EDM}}

Non-zero coefficients of the higher dimension operators in
Eq.~\eqref{eq:Lag:unit:rot} induce in general electric dipole
moments in nucleons, atoms, and molecular systems via contributions to
the hadronic Lagrangian in Eq.~\eqref{eq:LeffN}. Here we summarize the status
of the induced dipole moments for the systems used in our fit, namely,
the electron, neutron, and mercury EDMs.

In addition to these EDMs there are also constraints from the
experimental measurement of the muon EDM~\cite{Bennett_2009}, as well
as other systems with a hadronic component.  The direct constraint
from the current muon EDM measurement leads to constraints that are
about six orders of magnitude weaker than the one obtained via virtual
muon contributions to the electron EDM, and will thus not be used.
Concerning other hadronic EDMs, we have checked that within our
setting the experimental bounds on the radium~\cite{Bishof:2016uqx}
and xenon~\cite{PhysRevA.100.022505} EDMs are not competitive with the
neutron and mercury bounds and we do not include them in the fit
either.

\subsection{Electron}

The most recent experimental bound on the electron EDM is \cite{ACME:2018}
\begin{equation}
  |d_e| < 1.1 \times 10^{-29}\,e\,\text{cm} \qquad\text{@ 90\% confidence level (CL)}\,.
 \label{eq:eEDMexp}
\end{equation}
This value was obtained using ThO molecules, neglecting any CP-violating electron--nucleon
couplings. Bounds on the electron EDM have also been obtained using
YbF~\cite{Hudson2011} and HfF$^+$~\cite{PhysRevLett.119.153001}. The
resulting bounds are currently not competitive with the ThO bound and
are thus not used in our fit.

\subsection{Neutron}

The simplest hadronic system used in our fit is the neutron. The most
recent experimental bound on the neutron EDM is~\cite{Pignol:2021uuy}
\begin{equation}
 |d_n| < 1.8\times 10^{-26} e\,\text{cm}\qquad\text{@ 90\% CL} \,.
 \label{eq:nEDMexp}
\end{equation}
The future projections estimate an improvement of the limit 
to $|d_n| < 10^{-27}\,e\,$cm~\cite{Pignol:2021uuy}. 
Throughout this work, we assume that the $\theta_{\text{QCD}}$ term has negligible 
effect on any EDMs and all effects arise from the Yukawa SMEFT operators. 
The dipole moments of the
partons then contribute to the neutron EDM as
\begin{equation}\label{eq:dn}
  \frac{d_n}{e} = (1.1\pm 0.55)(\tilde d_d + 0.5\tilde d_u)
                  + \left(  g_T^u\frac{d_u}{e}
                          + g_T^d\frac{d_d}{e}
                          + g_T^s\frac{d_s}{e}\right) 
		\pm \frac{74}{3} (1\pm0.5) w \, \text{MeV} \,.
\end{equation}
The hadronic matrix elements of the chromoelectric dipole operators ($\tilde d_q$) 
and the Weinberg operator ($w$) are estimated using QCD sum rules and chiral
techniques~\cite{Pospelov:2005pr, Engel:2013lsa, Haisch:2019bml}. For
the matrix elements of the electric dipole operators ($d_q$) we use the lattice
results~\cite{Gupta:2018lvp} $g_T^u = -0.204(15)$, $g_T^d =
0.784(30)$, $g_T^s = -0.0027(16)$. The sign of the hadronic matrix
element of the Weinberg operator is not known; to be definite, we
choose the positive sign in our analysis. Note that in our scenario
the contribution of the Weinberg operator is always subdominant
(either suppressed by small quark masses, or numerically small compared
to the electron EDM bound), such that the sign ambiguity does not
affect the results of our global fit.

\subsection{Mercury}

Significant constraints in our fit arise also from the mercury EDM. 
The experimental bound is~\cite{Graner:2016ses}
\begin{equation}\label{eq:Hg:exp}
  |d_\text{Hg}| < 7.4 \times 10^{-30} e \,\text{cm} \qquad \text{@ 95\% CL} \,.
\end{equation}
The relation of the partonic EDMs to that of mercury is given
by~\cite{Engel:2013lsa}
\begin{equation}\label{eq:Hg:prediction}
  \frac{d_\text{Hg}}{e}
= \kappa_S
  \bigg[ g_{\pi NN} \bigg(   \frac{a_0}{e} \bar g_{\pi NN}^{(0)} 
                         + \frac{a_1}{e} \bar g_{\pi NN}^{(1)} \bigg) 
         + s_n \frac{d_n}{e} + s_p \frac{d_p}{e} \bigg]
  + a_e \frac{d_e}{e} \,.
\end{equation}
Here, $\kappa_S = - 2.8 \times 10^{-4}\,$fm$^{-2}$ denotes the
contribution of the Schiff moment to the mercury EDM, with an error
not exceeding 20\%~\cite{Dzuba:2002kg}. The expression in square
brackets is the Schiff moment for mercury. The CP-odd isoscalar and
isovector pion-nucleon interactions are given by $g_{\pi NN}^{(0)} =
(5 \pm 10) \times (\tilde d_u + \tilde d_d\big)\,$fm$^{-1}$, $g_{\pi
  NN}^{(1)} = 20_{-10}^{+40} \times (\tilde d_u - \tilde
d_d\big)\,$fm$^{-1}$, obtained from QCD sum-rule
estimates~\cite{Pospelov:2001ys}, while $g_{\pi NN} \simeq
13.5$~\cite{Stoks:1992ja, vanKolck:1996rm} is the CP-even pion-nucleon
coupling. The contribution of these interactions to the Schiff moment
is given by the parameters $a_0 = 0.01\,e\,$fm$^3$ and $a_1 = \pm
0.02\,e\,$fm$^3$. We took these ``best values'' from
Ref.~\cite{Engel:2013lsa}; they have an intrinsic uncertainty of about
an order of magnitude. The contributions of unpaired proton and
neutron to the Schiff moment has been calculated in
Ref.~\cite{Dmitriev:2003sc}, with result $s_p = 0.20(2)\,$fm$^2$ and
$s_n = 1.895(35)\,$fm$^2$. Finally, the contribution of the electron
EDM is subdominant; no explicit uncertainty is given for the prefactor
$a_e = 10^{-2}$~\cite{Pospelov:2001ys, Engel:2013lsa}; however,
different evaluations lead to different
signs~\cite{Ginges:2003qt}. See Ref.~\cite{Engel:2013lsa} for a
detailed discussion of all contributions to the mercury EDM.
Employing ``central values'' for all parameters, we find numerically
\begin{equation}\label{eq:Hg:prediction:num}
  \frac{d_\text{Hg}}{e}
= - 3.8 \times 10^{-4} 
  \bigg[ 0.5 \big(\tilde d_u + \tilde d_d\big)
	\pm 4 \big(\tilde d_u - \tilde d_d\big) \bigg]
  - 5.3 \times 10^{-4} \frac{d_n}{e} \,,
\end{equation}
with $d_n/e$ given in Eq.~\eqref{eq:dn}.

We have neglected several contributions in
Eq.~\eqref{eq:Hg:prediction}. The isotensor pion--nucleon coupling
contributes in principle, but is expected to be small in comparison to
the scalar and vector coupling, as it arises at higher order in chiral
perturbation theory~\cite{Engel:2013lsa}. Contributions of
four-fermion operators are smaller than the chromo EDM contributions by two
orders of magnitude, and are also neglected. Finally, in our analysis
we neglected the proton EDM contribution that is one order of
magnitude smaller than the neutron EDM contribution, as well as the
small electron EDM contribution.

\section{Collider observables\label{sec:collider}}

The SMEFT couplings in the Lagrangian in Eq.~\eqref{eq:Lag:unit:rot}
do not only induce electric dipole moments, but also affect the
production cross sections and decay branching ratios of the Higgs
boson. In this section we give an overview of the most important
effects; the actual numerical implementation of the LHC constraints is
discussed in Sec.~\ref{sec:global}.  The main effects are captured by
considering modifications of the gluon fusion production cross section
($gg\to h$) and the $h \to \gamma \gamma$ branching ratio (both are
one-loop induced both in the SM and in our setup), as well as of the
branching ratios to fermions observed by ATLAS and CMS ($h\to b\bar
b$, $h\to c\bar c$, $h\to\tau\tau$, $h\to\mu\mu$), and the total decay
width of the Higgs.  It is convenient to parameterise these
modifications in terms of the parameters
\begin{equation}\label{eq:def:rf}
  r_{f,\pm} \equiv \frac{v}{\sqrt{2}m_{f}}\frac{v^2}{\Lambda^2} C_{fH\pm}\,. 
\end{equation}
Using Eq.~\eqref{eq:kappa} we can readily translate $r_{f,\pm}$ to the
$\kappa$-framework parameters, i.e., $\kappa_f \cos\phi_f = 1 -
r_{f,+}$ and $\kappa_f\sin\phi_f = - r_{f,-}$.

The deviation from the SM gluon fusion cross section or the decay to
gluon-induced light jets can be effectively captured by (see
Ref.~\cite{Brod:2013cka} for details)
\begin{equation}
\begin{split}\label{eq:mug}
R_{gg} \equiv 
\frac{\sigma (gg \to h)}{\sigma (gg \to h)_{\rm SM}}
	= \bigg(   \Big|\sum_q \big(1-r_{q,+}\big) A(\tau_{q}) \Big|^2
                 + \Big| \tfrac{3}{2} \sum_q r_{q,-} B(\tau_{q}) \Big|^2\bigg)
          \bigg/ \Big| \sum\limits_i A(\tau_{q}) \Big|^2 \,,
\end{split}
\end{equation}
where we have defined $\tau_{q} \equiv 4\,m_{q}^2/M_h^2 -
i\varepsilon$, the sum runs over all quark flavours $q=u,d,s,c,b,t$,
and the fermionic one-loop functions are given by
\begin{equation}
A(\tau) = \frac{3\tau}{2}\,\Big( 1 + (1-\tau)
\arctan^2\frac{1}{\sqrt{\tau-1}} \Big) \,, \qquad B(\tau) = \tau
\arctan^2\frac{1}{\sqrt{\tau-1}} \,. 
\end{equation}
Similarly, the modification of the Higgs decays into two photons with
respect to the SM reads
\begin{equation}\begin{split}\label{eq:mugammagamma}
R_{\gamma\gamma} & \equiv \frac{\Gamma(h\to \gamma\gamma)}{\Gamma(h\to \gamma\gamma)_{\rm SM}} \\
	        & = \frac{  \Big|A_W + \frac16\sum_f Q_{f}^2n_c(f) \big(1-r_{f,+}\big) A(\tau_{f}) \Big|^2 
		          + \Big|\frac14\sum_f Q_{f}^2n_c(f) r_{f,-} B(\tau_{f}) \Big|^2}
                  {\Big|A_W + \frac16 \sum_f Q_{f}^2 n_c(f) A(\tau_{f})\Big|^2}\,,
\end{split}
\end{equation}
where the sums run over all charged fermions $f$ with $n_c(f)=3$ for
quarks and $n_c(f) = 1$ for leptons, and the bosonic contribution is
given by
\begin{equation}
A_W = - \frac{1}{8} \left ( 2 + 3 \, \tau_W +
      3 \, \tau_W  \left (2-\tau_W \right )
    \arctan^2\frac{1}{\sqrt{\tau_W-1}} \right ) \,,
\end{equation}
with $\tau_W = 4M_{W}^2/M_h^2 - i\varepsilon$.  We keep the
modifications of the Higgs coupling to $WW$ and $ZZ$ bosons unmodified
as the contributions from Yukawa operators are loop induced compared
to the tree-level contributions already present at the dimension-four
level.  Moreover, $h\rightarrow Z\gamma$ is omitted as its decay width
is suppressed by a smaller coupling compared to
$h\rightarrow\gamma\gamma$.

The signal strengths of searches for Higgs decays to fermions are also affected
by the corresponding modifications of the partial widths, namely:
\begin{equation}\label{eq:Rff}
  R_{ff} \equiv \frac{\Gamma(h\to ff)}{\Gamma(h\to ff)_{\rm SM}} = (1-r_{f,+})^2 + r_{f,-}^2 \,.
\end{equation}
As of now, the ATLAS and CMS collaborations have observed Higgs decays
into $f=b,c,\tau,\mu$.  However, the modifications of the partial
widths to any fermion affect the total width of the Higgs via
\begin{equation}
    \frac{\Gamma^{\text{tot}}}  {\Gamma^{\text{tot}}_{\text{SM}}} = 1 
    +\sum\limits_{f}                (R_{ff}-1)\mathrm{Br}(h\rightarrow \bar f f)_\text{SM} 
    +\sum\limits_{X=gg,\gamma\gamma}(R_{X} -1)\mathrm{Br}(h\rightarrow        X)_\text{SM} \,,
	\label{eq:totalrate}
\end{equation}
which in turn affects all its branching ratios, i.e., all the signal
strength for Higgs searches.

We conclude this section with some remarks concerning the convergence
of the SMEFT expansion and the potential impact of dimension-eight
operators. The issue arises because the squared amplitude enters the
expression~\eqref{eq:Rff}, leading to terms proportional to
$1/\Lambda^4$ in $R_{ff}$ that we keep in our numerics. What would be
the impact of including dimension-eight Yukawa operators (of the
generic form $Q^{(8)} \sim (H^\dagger H)^2 \bar{Q}_L H C_{qH} q_R$) in
the amplitude? While this question is hard to answer quantitatively
without actually performing the analysis, the following arguments
suggest that the impact would result only in minimal changes on the
bounds of the dimension-six coefficients.

First, note that this issue concerns only the real parts of the Wilson
coefficients. As there is no interference with the SM in the CP-odd
part of the amplitude, such dimension-eight terms would be of order
$1/\Lambda^8$. The first term on the right side of Eq.~\eqref{eq:Rff},
on the other hand, would be changed to
\begin{equation}\label{eq:rf:8}
  (1-r_{f,+})^2 \to
1 - 2 C^{(6)} \frac{v}{\sqrt{2}m_f} \frac{v^2}{\Lambda^2}
  + \big(C^{(6)}\big)^2 \frac{v^2}{2m_f^2} \frac{v^4}{\Lambda^4}
  - C^{(8)} \frac{v}{\sqrt{2}m_f} \frac{v^4}{\Lambda^4}
  + {\mathcal O}\big(\Lambda^{-6}\big) \,.
\end{equation}
Here, $C^{(6)}$ and $C^{(8)}$ denote generic dimension-six and
dimension-eight Wilson coefficients, respectively. We see that the
additional last term in Eq.~\eqref{eq:rf:8} is parametrically
suppressed with respect to the linear dimension-six term by a factor
$v^2/\Lambda^2$, and with respect to the quadratic dimension-six term
by a factor $m_f/v$. A potential problem arises if the term quadratic
in $C^{(6)}$ dominates the fit. A direct comparison between the last
two terms in Eq.~\eqref{eq:rf:8}, under the ``EFT assumption''
$C^{(6)} \sim C^{(8)}$, shows that the dimension-eight contribution
will be subleading as long as $C^{(6)} > \sqrt{2}m_f/v$. We will see
in Fig.~\ref{fig:1flavour} that this condition is clearly satisfied
for all fermions apart from the top quark. While this is only a rough
estimate, we interpret this as an indication that the impact of the
dimension-eight terms on the fit would be tiny. By contrast,
Fig.~\ref{fig:1flavour_3rdgen} shows that for the top quark $C_{tH+}
\lesssim 3$, which is larger than $\sqrt{2}m_t/v \sim 1$, such that
for values of $C^{(8)} \sim 9$ the dimension-eight term would
contribute significantly to the fit. While this required value of
$C^{(8)}$ is close to the perturbativity limit, we conclude that the
EFT expansion in the collider observables might not work as well for
the top quark as it does for all other fermions.
CP-odd contributions, on the other hand, are not expected to receive
large corrections from dimension-eight operators for any of the
fermions, as discussed above.

\section{Global Analyses\label{sec:global}}

Having obtained in the previous sections the expressions for the relevant observables, EDMs, 
Higgs production cross sections ($\sigma$), and branching ratios (BR) decay widths we now 
combine the constraints on the considered Wilson coefficients in a global analysis. 
For this purpose we use the {\small\texttt{GAMBIT}} global fitting framework~\cite{GAMBIT:2017yxo} to
calculate a combined likelihood based on these two sets of
constraints. The collider likelihoods are taken from the
{\small\texttt{HiggsSignals\_2.5.0}} and {\small\texttt{HiggsBounds\_5.8.0}}
codes~\cite{Bechtle:2020uwn, Bechtle:2020pkv} interfaced with the  
{\small\texttt{ColliderBit}} module \cite{GAMBIT:2017qxg} of  {\small\texttt{GAMBIT}}.

{\small\texttt{HiggsSignals}} contains three modules to compute likelihoods from Higgs 
measurement at the LHC. Each module calls {\small\texttt{HiggsBounds}} to calculate 
the various production cross sections and branching ratios within our SMEFT model and 
to obtain the signal strengths implemented in each module.

The first module computes a likelihood of Run~1 Higgs measurements using 
a set of signal strengths provided by the ATLAS and CMS combination of 
Run~1 data \cite{ATLAS:2016neq}.
In this module, the signal strengths are ``pure channels'' in the sense that one decay channel is combined 
with one specific production channel, i.e., 
$\mu_i^f = (\sigma_i \text{BR}^f)/(\sigma_{i,\text{SM}} \text{BR}^f_{\text{SM}})$
with $i$ and $f$ indicating different production modes ($ggh$, VBF, $Wh$, $Zh$, $tth$)
and decay channels ($ZZ$, $WW$, $\gamma\gamma$ $\tau\tau$, $b\bar b$), respectively.
Run 1 data are sensitive to $20$ such signal strengths.
The (Log)likelihood is then obtained from
\begin{align}
\chi^2
= ({\boldsymbol{\mu}}-\boldsymbol\mu^\text{exp})^TC_\text{cov}^{-1}(\boldsymbol\mu-\boldsymbol\mu^\text{exp}) \,,
\label{eq:chi2HS}
\end{align}
where $\boldsymbol\mu$ are the vectors containing the $20$ signal strengths 
computed as a function of the SMEFT Wilson coefficients, 
$\boldsymbol\mu^\text{exp}$ are the corresponding experimental combinations, 
and the superscript ``$T$'' denotes transposition. 
The matrix $C_\text{cov}$ is the signal-strength covariance matrix describing 
the uncertainties of and correlations among the signal strengths. 

For Run~2 measurements there are two {\small\texttt{HiggsSignals}} modules
to compute a likelihood, one that uses the simplified template cross section 
measurements (STXS)~\cite{LHCHiggsCrossSectionWorkingGroup:2016ypw} and another
that uses the peak-centered method. In our analysis we mainly use the former
since the peak-centered method increases the computing time by almost an
order of magnitude and has also less constraining power, as we have checked explicitly. 
The only exception in which we do use the peak-centered-method module is to 
include the two $h\to\mu\mu$ analyses containing in total $34$ peak measurements
\cite{ATLAS:2020fzp,CMS:2020xwi}.
Each signal strength is a combination of various production modes 
weighted by the corresponding experimental efficiency ($\epsilon_i$), which 
accounts for the detector performance in identifying signal events, i.e., 
\begin{align}
\mu = \frac{\sum_j^{N} (\sigma\cdot \text{BR})_j \epsilon_j }
{\sum_j^{N} (\sigma^\text{SM}\cdot \text{BR}^\text{SM})_j\epsilon^\text{SM}_j} \,.
\end{align}
In our analysis we include the signal strengths of $56$ measurements originating 
from experimental searches from June 2021 \cite{ ATLAS:2018jvf,
  ATLAS:2017fak, ATLAS:2018xbv, ATLAS:2018ynr, ATLAS-CONF-2019-045,
  CMS:2017odg, CMS:2017bcq, CMS:2018hnq, ATLAS:2019vrd, ATLAS:2020fcp,
  CMS:2018nak, CMS:2018fdh, CMS-PAS-HIG-18-030, CMS-PAS-HIG-18-019,
  CMS-PAS-HIG-19-001, CMS-PAS-HIG-18-029, CMS-PAS-HIG-18-032,
  ATLAS-CONF-2019-029, ATLAS:2020rej, CMS:2020dvg}.
The corresponding (Log)likelihood is then obtained in an analogous manner as
for the Run 1 data in Eq.~\eqref{eq:chi2HS}. 
The SM predictions
for the cross sections are obtained through a fit to the predictions
from Yellow Report~4~\cite{LHCHiggsCrossSectionWorkingGroup:2016ypw}.

To compute a likelihood from the EDM measurements, we assume their 
experimental uncertainties $\sigma_\text{exp}$ to be Gaussian distributed yielding  
the likelihoods ($\log \mathcal{L}$)
\begin{align}
\chi^2_X
 = -2\log\mathcal{L}_X
 = \frac{\big(d_X(2\,\text{GeV}) - d_X^\text{\,exp}\big)^2}{\sigma_{X,\text{exp}}^2}\,,
\label{eq:chi2}
\end{align}
with $X=n,\,e,\,\text{Hg}$. The uncertainties $\sigma_{X,\text{exp}}$ are obtained
from the upper limits given in Eqs.~\eqref{eq:eEDMexp}, \eqref{eq:nEDMexp}, and~\eqref{eq:Hg:exp}
by assuming zero central values for $d_X^\text{\,exp}$.

The total likelihood of a parameter point is computed from the  sum of the $\chi^2$ values 
from the EDM and LHC likelihoods. To identify preferred regions in a subspace of the 
scanned parameters we must profile over the remaining scanned parameters. We perform the
profiling using {\small\texttt{pippi}}~\cite{Scott_2012}, which computes the 
lowest possible $\chi^2$ value of each parameter point in the subspace of interest
by allowing all other parameters to float simultaneously to minimize the $\chi^2$.
When projecting onto a two-parameter subspace, the allowed regions at 
68\% and 95\% CL correspond to the parameter space for which the difference 
$\chi^2 - \chi^2_{\text{best}}$ is less or equal than $\chi^2_\text{68\%}\approx2.28$ and 
$\chi^2_\text{95\%}\approx5.99$, respectively.
$\chi^2_{\text{best}}$ is the $\chi^2$ value of the best-fit point, i.e., the lowest $\chi^2$ value.
In our case the $\chi^2$ value of the SM (all SMEFT Wilson coefficients are zero)
divided by the number of the included $N_\text{d.o.f.}=79$ observables 
is $\frac{\chi^2_\text{SM}}{N_\text{d.o.f.}} = 0.83$,
where we did not include the $34$ peak observables from the $h\to\mu\mu$ measurements;
they only affect the muon Yukawa.
In the scan that we present the difference between the $\chi^2$ of the best-fit point and the SM
is always less than $0.1$, so we do not display the best-fit value in the plots.

\subsection{One-flavour scans}

\begin{figure}[ph]
	\includegraphics[]{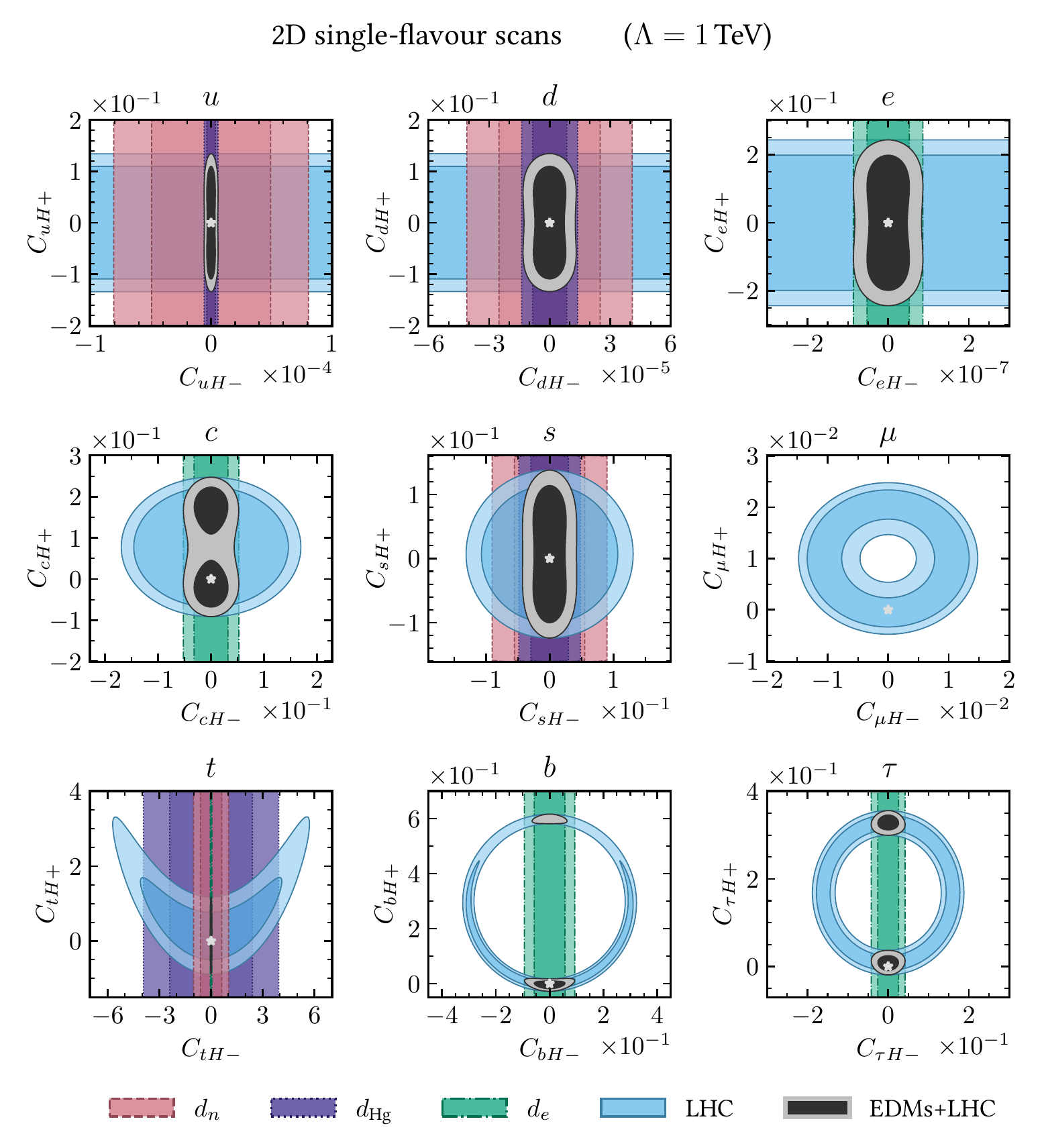}
	\caption{
      EDM and LHC constraints on CP-even ($C_{fH+}$) and CP-odd ($C_{fH-}$) Higgs Yukawa 
      couplings assuming $\Lambda=1$\,TeV.  Each plot shows the result of a 2D scan in which
      only the two Wilson coefficients of the respective fermion are sampled. 
      The contours represent the allowed 68\% and 95\%
      confidence regions for their respective observables.
      The colour coding of individual constraints is given in the legend, 
      gray/black areas correspond to the regions allowed by the combination of all constraints.
      Regions allowed by an individual observable are only shown if 
      the observable has a constraining effect on the
      parameter space.
      The sampled intervals are bounded by the perturbativity condition $|C_{fH\pm}| \leq 4\pi$.
    \label{fig:1flavour}}
\end{figure}

To set the stage, we first perform two-dimensional scans over the
Wilson coefficients of each fermion individually, with the Wilson
coefficients of all other fermions set to zero. The results are shown
in Fig.~\ref{fig:1flavour}. We allow the Wilson coefficients to float
within their perturbative values, $|C_{fH\pm}| \leq 4\pi$, and fix
$\Lambda = 1\,$TeV in our scan. All numerical input parameters are
taken from Ref.~\cite{Zyla:2020zbs}.

In this work, we parameterize the effects beyond the SM by the SMEFT
Wilson coefficients in Eq.~\eqref{eq:Lag:unit:rot}. EDMs only
constrain the imaginary part of the Wilson coefficients, $C_{fH-}$,
while LHC observables constrain also the real parts, $C_{fH+}$. In
general, the electron EDM gives the strongest bounds on CP-violating
parts of the lepton and heavy-quark (top, bottom, charm) coefficients,
while the corresponding light-quark (up, down, strange) coefficients
are mainly constrained by a combination of neutron and mercury EDM
measurements.

\bigskip

\begin{figure}[t]
	\center
    % if you want to shrink the plot do it in python not here
	\includegraphics[]{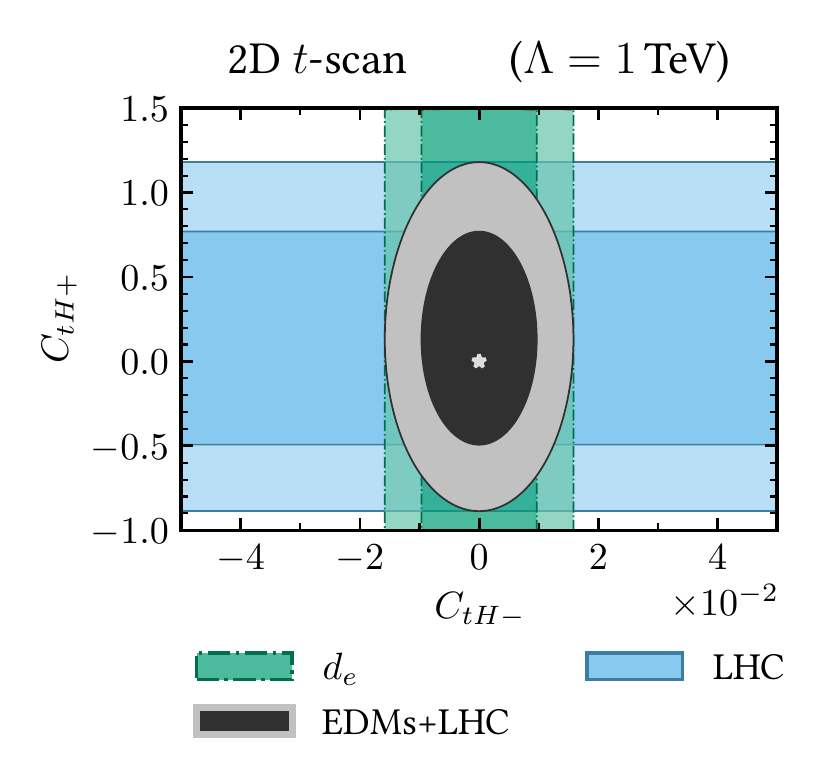}
	\caption{Magnification of the lower left plot of Fig.~\ref{fig:1flavour} showing the 
      constraints on CP-even and CP-odd contributions to the top-quark Yukawa ($C_{tH\pm}$).
    \label{fig:1flavour_3rdgen}}
\end{figure}

$\bullet$ {\bf Top}\\ 
\noindent
The constraints from the modification of Higgs
production in gluon fusion have the parametric dependence $\mu_{gg} =
(1-r_{t,+})^2 + r_{t,-}^2$, where we used the asymptotic values
$A(\infty) = B(\infty) = 1$, valid to good approximation for the top
quark, and neglected the contribution of all other quark flavours. In
the same approximation, we have $\mu_{\gamma \gamma} = (1.0 +
0.27r_{t,+})^2 + 0.17r_{t,-}^2$. These are the dominant LHC
constraints on the top couplings; note, however, that all production
channels are included in our analysis. The corresponding constraints
are shown in Fig.~\ref{fig:1flavour} (bottom left panel). The strongest
individual bound on the CP-odd coefficient $C_{tH-}$ arises from the
electron EDM, while the constraints arising from the hadronic systems
are much weaker, see Fig.~\ref{fig:1flavour}. A zoomed-in version of
the combined fit region is shown in Fig.~\ref{fig:1flavour_3rdgen}.

\bigskip

$\bullet$ {\bf Bottom, charm, tau, muon}\\ 
\noindent
None of the bottom, charm, tau,
and muon EDMs themselves have been measured with high
precision. Consequently, EDM bounds on CP-odd Higgs couplings of the
fermions arise from their virtual contributions to the electron EDM
(again, the bounds arising from hadronic systems are negligible in
comparison). It is remarkable that, the presence of a large quadratic
logarithm (see Eq.~\eqref{eq:L3}), makes the electron EDM bound weaker
than that on the top by only a factor~$5$ for the bottom, a factor~$2$
for the charm, a factor~$6$ for the muon, and of the same order for
the tau, rather than being suppressed by the much larger factor $Q_t^2
m_t/ Q_f^2 m_f = {\mathcal O}(50 - 500)$, expected
naively~\cite{Panico:2018hal} from the Yukawa suppression.  For the
bottom and charm quarks, this strong logarithmic enhancement indicates
that QCD corrections are large in these cases and should be
included~\cite{BS2022}, similar to the case of hadronic dipole
moments~\cite{Brod:2013cka, Brod:2018pli} (see the discussion in
Sec.~\ref{sec:EFT:below}).

Measurements at the LHC directly constrain the decay of the Higgs into
bottom, charm, tau, and muon pairs.  The main effect comes from the
modification of the partial $h\to f\bar f$ widths (see
Eq.~\eqref{eq:Rff}).  This is a simplified picture, as modifying any
Yukawa also affects the total Higgs decay width (see
Eq.~\eqref{eq:totalrate}), the $h \to \gamma \gamma$ decay and, for
quarks, also Higgs production via gluon-fusion. This effect is largest
for the bottom quark (that dominates the SM Higgs decay width);
nevertheless, we include the effect of all fermions on the total Higgs
width.

Note that a negative value for the bottom Yukawa (in the sense of
Eq.~\eqref{eq:kappa}) is excluded at $68\%\,$CL (bottom middle panel
in Fig.~\ref{fig:1flavour}).  Note also that for the muon Yukawa, the
recent LHC measurements are more constraining than the electron EDM by
an order of magnitude.  Regarding tau couplings, the CMS analysis on
the CP structure of the $h \to \tau \tau$ decay~\cite{CMS:2021sdq}
disfavours large values of $|C_{\tau H-}|$.  While this analysis is
not included here, as it is not (yet) part of the
{\small\texttt{HiggsSignals}} data set, it would split the 2$\sigma$
LHC constraint (blue region in lower right panel of
Fig. \ref{fig:1flavour}) in two distinct regions as illustrated in the
CMS analysis \cite{CMS:2021sdq}.

\bigskip

$\bullet$ {\bf Electron, up, down, strange}\\ 
\noindent
The main EDM bounds on the
electron and light-quark (up, down, strange) SMEFT coefficients arise
from their contributions to the electron EDM and the neutron and
mercury EDMs, respectively, as discussed in Sec.~\ref{sec:match}. Note
the different impact of the neutron and mercury EDMs on the up and
down coefficients, due to the strong isospin dependence of the
corresponding EDM predictions.

LHC bounds on the Wilson coefficients for the electron and the up and
down quarks arise from modifications of the total Higgs decay width
Eq.~\eqref{eq:totalrate}. Indeed, we checked that the contributions to
gluon fusion and $h \to \gamma \gamma$ are subleading.  (For
modifications of Higgs production induced by parton distribution
functions, see Ref.~\cite{Soreq:2016rae}.)  However, the resulting
constraints are very weak when compared to the SM Yukawas, as
illustrated by the corresponding ratios $|r_{e,+}| \lesssim 4000$,
$|r_{u,+}| \lesssim 500$, and $|r_{d,+}| \lesssim 225$. In contrast,
the bound on the CP-odd strange-quark coefficient from LHC is almost
competitive with the hadronic EDM bounds (central panel in
Fig.~\ref{fig:1flavour}).

\subsection{Two-, three-flavour scans, and beyond}

Next, we let the Wilson coefficients of more than one fermion flavour
float simultaneously. This allows for the cancellation of the
contributions of the considered Wilson coefficients to the
constraining EDMs, thereby relaxing the bounds in certain regions of
parameter space. There are no such cancellations possible in the
collider observables, because the main effect comes from partial decay
widths where no interference is possible. (The small interference term
between top- and bottom-quark contribution to Higgs production via
gluon fusion and $h \to \gamma \gamma$ is negligibly small.) However,
the bounds on the different Wilson coefficients are still correlated.
For instance, the Higgs decay into bottom quark dominates the total
Higgs decay width, and thus the $h\to b \bar b$ rate affects {\em all}
branching ratios significantly.  The same is true, albeit to a lesser
extent, for all other Higgs decays.

\bigskip

\begin{figure}[t]
	\center
    % if you want to shrink the plot do it in python not here
	\includegraphics[]{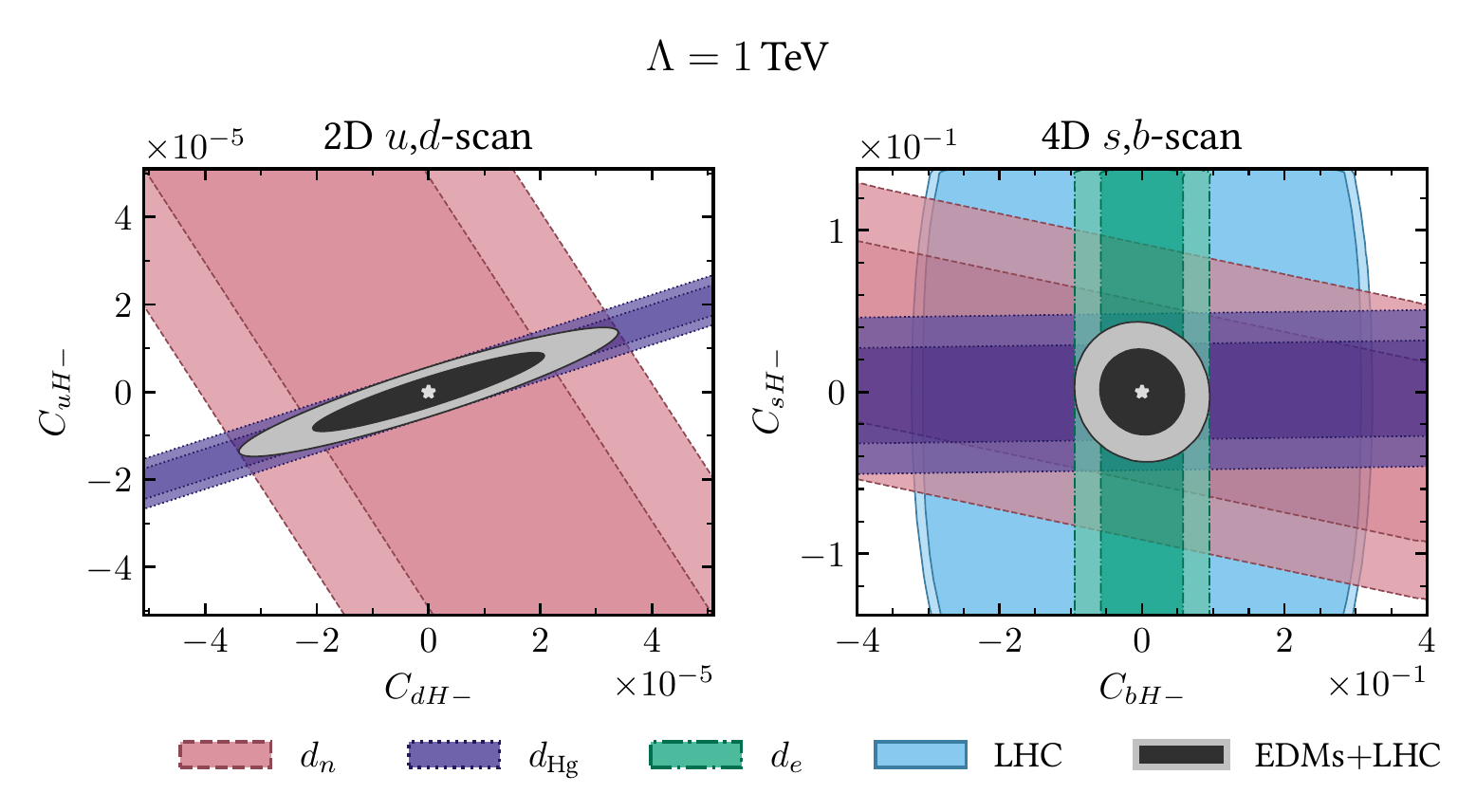}
	\caption{
          {\bf \itshape Left:} 
          Constraints resulting from the combined 2D scan of
          CP-odd up and down Yukawas ($C_{uH-}$, $C_{dH-}$).
          As the EDMs are not sensitive to the 
          CP-even parts and the corresponding LHC constraints are weak we do not scan over those parameters.
		  {\bf \itshape Right:} 
          Constraints resulting from the combined 4D scan of
          CP-even and CP-odd bottom and strange Yukawas ($C_{bH\pm}$, $C_{sH\pm}$).
          Contours represent the allowed 68\% and 95\% confidence regions,
          the colour coding of individual constraints is given in the legend, and 
          gray/black areas correspond to the combined regions.
          For more details see caption of Fig.~\ref{fig:1flavour}.
	    \label{fig:udbsscan}}
\end{figure}

$\bullet$ {\bf Up and down}\\
\noindent
In the left panel of Fig.~\ref{fig:udbsscan} we show the results of a
two-parameter scan over $C_{uH-}$ and $C_{dH-}$.  We do not scan over
the corresponding CP-even Wilson coefficients as they do not enter the
EDM predictions.  The different dependence of the neutron and the
mercury EDMs on the up- and down-quark coefficients is clearly
visible. With the isoscalar pion-nucleon coupling (entering the
mercury EDM prediction) being subdominant, the bands of the two
observables are nearly orthogonal, thus allowing to set stringent
constraints on both $C_{uH-}$ and $C_{dH-}$.

\bigskip

\newpage

$\bullet$ {\bf Bottom and strange}\\
\noindent 
In the right panel of Fig.~\ref{fig:udbsscan} we show the results of a
four-parameter scan over $C_{bH\pm}$ and $C_{sH\pm}$ after profiling
over the CP-even couplings.  As we neglect the tiny strange-quark
contribution to the electron EDM, the only constraint on $C_{sH-}$
arises from the two hadronic EDMs. On the other hand, $C_{bH-}$
receives its dominant constraint from the electron EDM, while the
contributions to the hadronic EDMs are also taken into account.

We include the CP-even Wilson coefficients in the scan as they are
both bounded by LHC measurements. As there is no direct measurement of
$h\to s\bar s$, the correlation between $C_{sH-}$ and $C_{bH-}$
results from the contributions to the total Higgs decay width. The
analogous plots with CP-even Wilson coefficients do not contain any
additional information compared to the one-flavour scans and are thus
not shown here.

\bigskip

\begin{figure}[ph]
    % if you want to shrink the plot do it in python not here
  \hspace*{-2em}\includegraphics[]{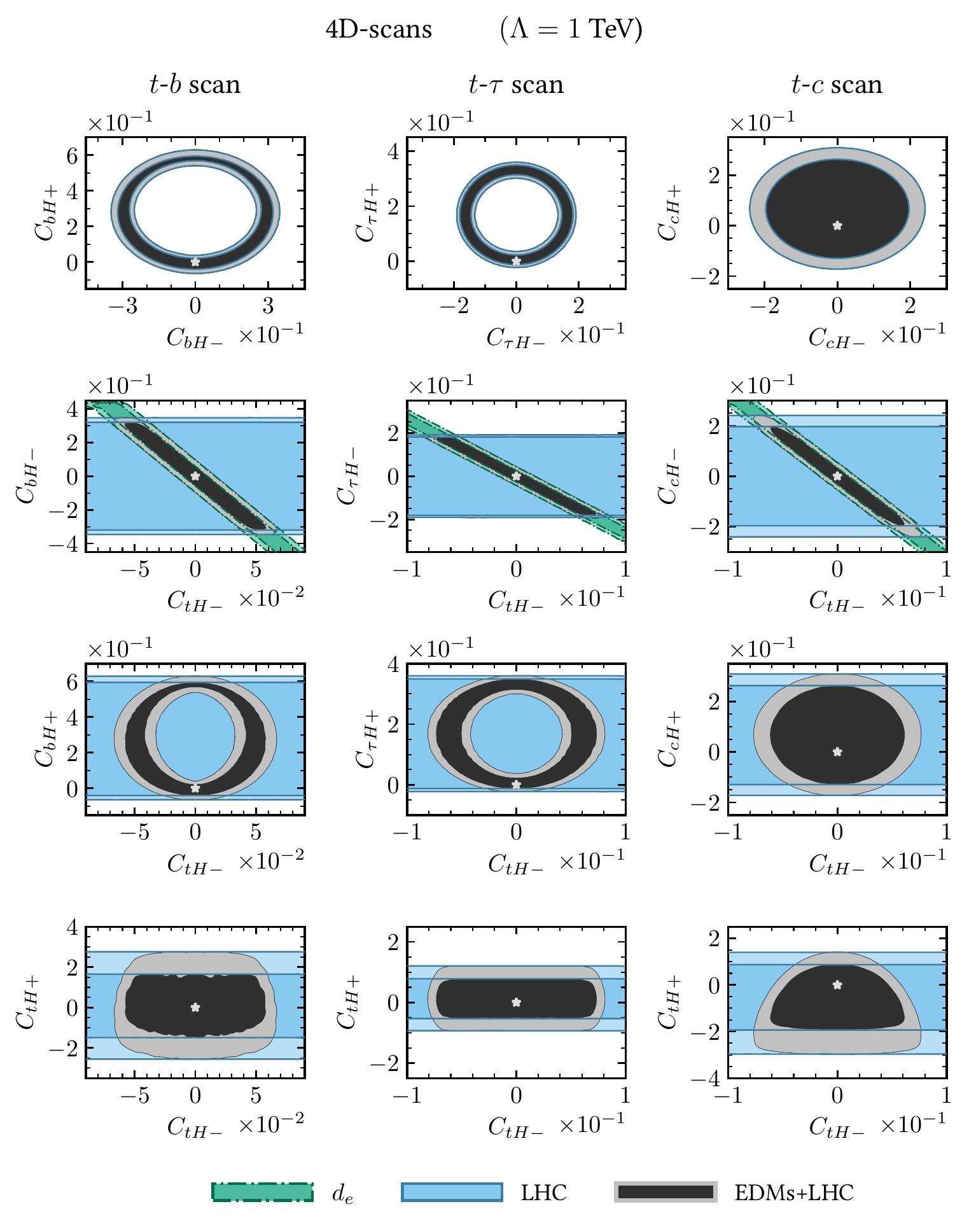}
	\caption{Constraints resulting from a 4D scan of top and
          bottom coefficients ($C_{tH\pm}$, $C_{bH\pm}$; first
          column), top and tau coefficients ($C_{tH\pm}$, $C_{\tau
            H\pm}$; second column), and top and charm coefficients
          ($C_{tH\pm}$, $C_{cH\pm}$; third column), assuming
          $\Lambda=1$\,TeV.  In each plot only two parameters are
          shown, the remaining two are profiled over (see main text).
          Contours represent the allowed 68\% and 95\% confidence
          regions, the colour coding of individual constraints is
          given in the legend, and gray/black areas correspond to the
          combined regions. For details see main text and the caption
          of Fig.~\ref{fig:1flavour}.
      \label{fig:tbtauscan}}
\end{figure}

$\bullet$ {\bf Top and bottom; top and tau; top and charm}\\
\noindent
In Fig.~\ref{fig:tbtauscan} we show the results of three different
four-parameter scans, floating $C_{tH\pm}$ simultaneously with
$C_{bH\pm}$ (first column), with $C_{\tau H\pm}$ (second column), and
with $C_{cH\pm}$ (third column). The electron EDM bounds on $C_{tH-}$
could, in principle, be lifted by two orders of magnitude compared to
the single-flavour scan by choosing values close to the perturbativity
limit for $C_{bH-}$, $C_{\tau H-}$, and $C_{cH-}$. However, given the
LHC bounds on these parameters the bound on $C_{tH-}$ is weakened by
only a factor of the order of five.  Note also that allowing $C_{tH+}$
to float significantly increases the allowed range of values for
$C_{bH+}$ at the $68\%\,$CL (upper left panel in
Fig.~\ref{fig:tbtauscan}). The relaxation of the bounds compared to
the single-flavour fit is even more pronounced for the charm quark
(upper right panel in Fig.~\ref{fig:tbtauscan}). In the last row, we
present the constraints on the parameters $C_{tH+}$ and $C_{tH-}$,
with the other parameters profiled. This can be directly compared to
Fig.~\ref{fig:1flavour_3rdgen}, showing a significant relaxation of
the bounds.

Interestingly, the combination of the electron EDM with LHC
constraints has a profound impact on the constraint on the CP-even
Wilson coefficients and not only on the CP-odd ones as one would
naively expect.  To understand this better, we first consider the
second panel from above in the first column of
Fig.~\ref{fig:tbtauscan} showing the allowed $C_{tH-}$--$C_{bH-}$
region.  Here, the allowed combined region corresponds to the
intersection of the electron EDM and the LHC constraint, which implies
that for {\em each} allowed pair of $(C_{tH-}, C_{bH-})$, it is
possible to find corresponding allowed values of $C_{bH+}$ and
$C_{tH+}$ (which have been profiled out in the plot). This can easily
be verified by looking at the bottom-left and bottom-center panels of
Fig.~\ref{fig:1flavour}.
By contrast, the electron EDM on its own does not constrain the
$C_{tH-}$--$C_{bH+}$ subspace at all, i.e., the whole parameter space
of the third panel in the first column of Fig.~\ref{fig:tbtauscan}
($C_{tH-}$--$C_{bH+}$ plot) is allowed with respect to the electron
EDM. The reason being that one can always cancel the top against the
bottom contributions to the EDM (see green band in the upper left
panel).  However, the combined EDM--LHC region is smaller than the one
allowed by LHC alone, as not all values of $C_{bH-}$ required to
cancel the contributions of $C_{tH-}$ are allowed by LHC bounds.  In
fact, the bottom single-flavour analysis (bottom center panel of
Fig.~\ref{fig:1flavour}) indicates that roughly $C_{bH-} \approx
C_{bH+}$, resulting in the much smaller allowed combined region in the
4D scan.  Similar arguments apply to the plots that show the top--tau
and top--charm coefficients. Regarding the case of the charm quark,
note that its contribution to the electron EDM is larger than that of
the bottom quark, while the LHC constraint is comparatively weaker,
resulting in a larger allowed combined region.  Finally, we remark
that while the contribution to the electron EDM of the muon is similar
to that of the bottom, the muon Yukawa is so strongly constrained by
recent LHC measurement that no appreciable cancellation can occur,
which is why we do not present this scan.

\bigskip
\newpage

\begin{figure}[hp]
    % if you want to shrink the plot do it in python not here
	\includegraphics[]{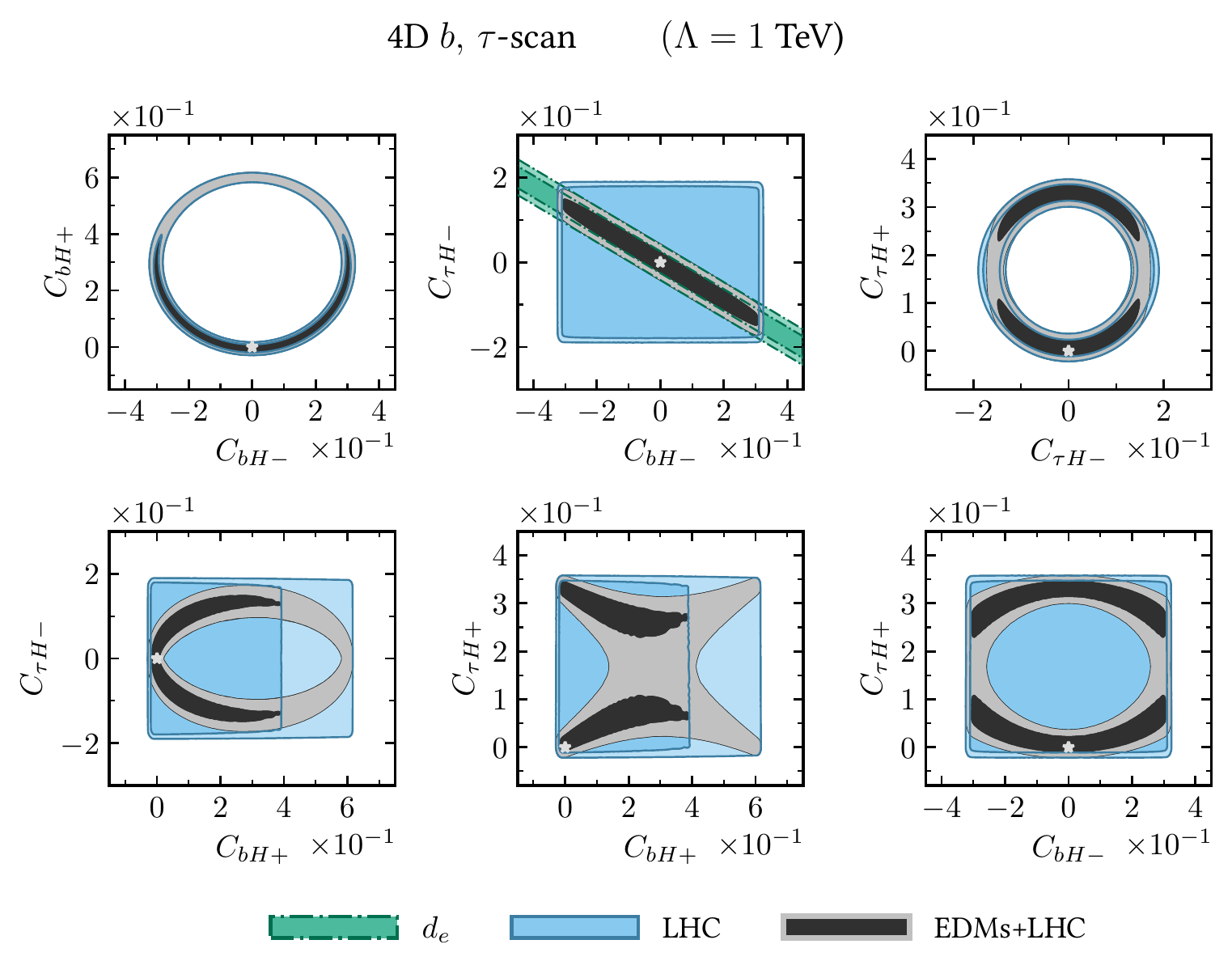}
	\caption{
      Constraints resulting from a 4D scan of bottom and $\tau$ Wilson 
      coefficients  ($C_{bH\pm}$, $C_{\tau H\pm}$) assuming $\Lambda=1$\,TeV.
      In each plot only two parameters are shown, the remaining two 
      are profiled over (see main text).
      Contours represent the allowed 68\% and 95\% confidence regions,
      the colour coding of individual constraints is given in the legend, and 
      gray/black areas correspond to the combined regions.
      For details see main text and the caption of Fig.~\ref{fig:1flavour}.
	\label{fig:btauscan}}
\end{figure}

$\bullet$ {\bf Bottom and tau}\\
\noindent
In Fig.~\ref{fig:btauscan} we show the results of a scan of the four
parameters $C_{bH\pm}$ and $C_{\tau H\pm}$.  As in the previous
four-parameter scans, there is an interesting interplay between EDM
and LHC bounds. When considering EDM bounds only, we can always cancel
the constraint if either $C_{bH-}$ or $C_{\tau H-}$ is
profiled. Hence, there are no pure EDM constraints in any but the
upper center panel where the CP-odd coefficients $C_{bH-}$ and
$C_{\tau H-}$ are displayed.  In contrast, the EDM constraints cannot
always be satisfied if also LHC constraints are included. In the upper
right plot with the two tau coefficients displayed, one can see that
the allowed, combined region is enlarged compared to the
single-flavour scan. However, for extreme values for $C_{\tau H-}$,
still allowed by LHC bounds, $C_{bH-}$ cannot be profiled such that it
compensates the large tau contribution to the electron EDM and
simultaneously still be within the 2$\sigma$-level bottom LHC
bounds. In contrast, $C_{\tau H-}$ can be profiled such that the $C_{b
  H-}$ coefficient in the upper left plot is only bounded by LHC
constraints. The reason for this is apparent from
Eq.~\eqref{eq:C3eew}: the contribution of the tau to the electron and
quark EDMs are larger than those of the bottom by about a factor $3
m_\tau \log^2(m_\tau/M_h)/ m_b \log^2(m_b/M_h) \approx 2$, meaning
that the bottom contribution can always be fully canceled by $\tau$
contributions, but not vice versa.  This implies that, given current
data, the combined bounds (apart from the combination $(C_{b H+}, C_{b
  H-})$) are more stringent than either the LHC bounds or the EDM
bounds alone. The effect of the electron EDM, further restricting the
parameter regions allowed by LHC data, is also clearly visible in the
combined region in the lower center panel that shows the bounds on the
two CP-even coefficients $C_{b H+}$ and $C_{\tau H+}$.

\bigskip

\begin{figure}[h]
	\center
    % if you want to shrink the plot do it in python not here
	\includegraphics[]{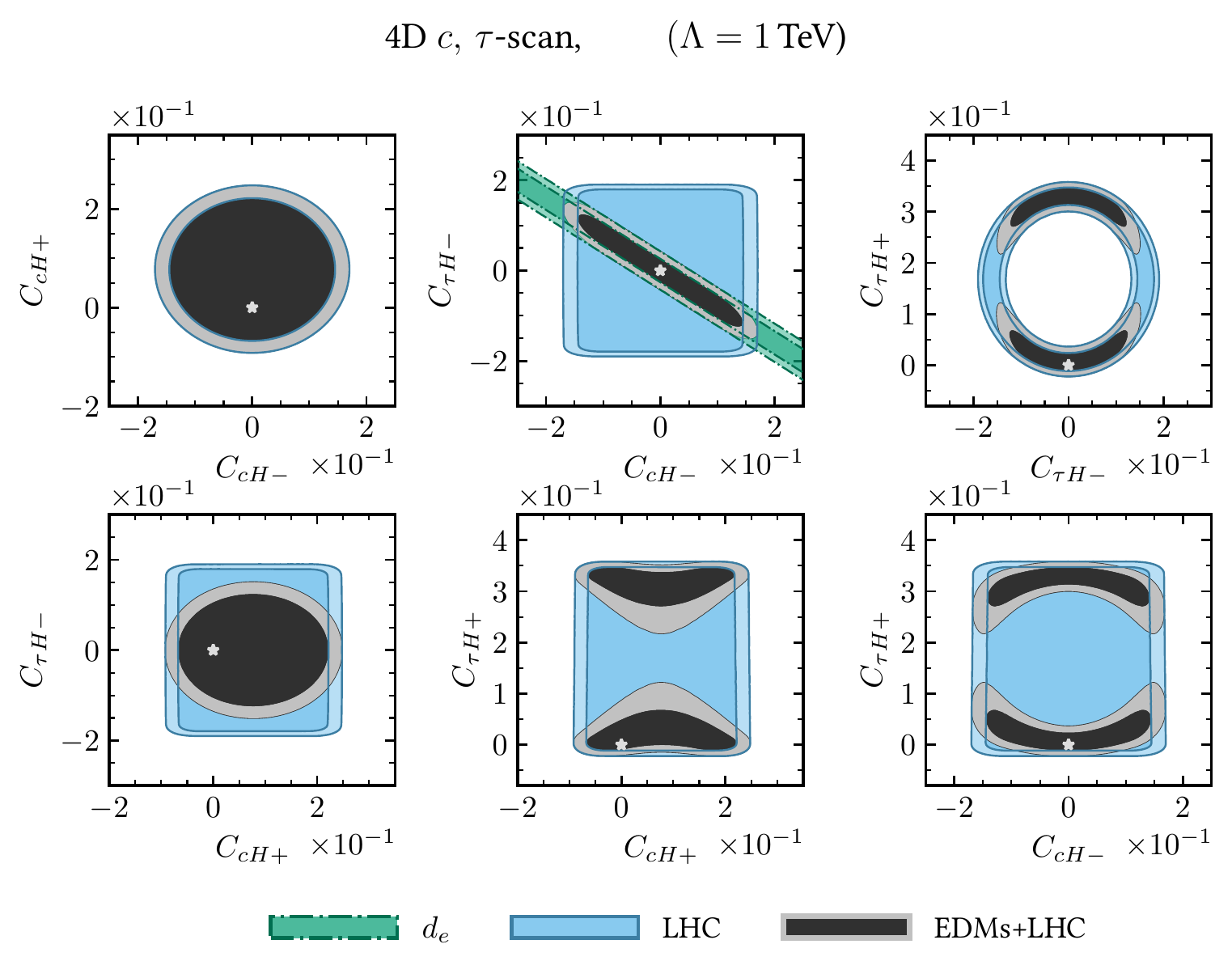}
    \caption{
      Constraints resulting from a 4D scan of charm and $\tau$ Wilson 
      coefficients  ($C_{cH\pm}$, $C_{\tau H\pm}$) assuming $\Lambda=1$\,TeV.
      In each plot only two parameters are shown, the remaining two 
      are profiled over (see main text).
      Contours represent the allowed 68\% and 95\% confidence regions,
      the colour coding of individual constraints is given in the legend, and 
      gray/black areas correspond to the combined regions.
      For details see main text and the caption of Fig.~\ref{fig:1flavour}.
    \label{fig:tctauscan}}
\end{figure}

$\bullet$ {\bf Charm and tau}\\
\noindent
In Fig. \ref{fig:tctauscan} we present the results of a scan of the
four parameters $C_{cH\pm}$ and $C_{\tau H\pm}$. The results are
analogous to the case of bottom and tau discussed above. Note that the
contribution to the electron EDM of the charm quark is larger than
that of the bottom quark by a factor of roughly $4 m_c
\log^2(m_c/M_h)/ m_b \log^2(m_b/M_h) \approx 1.6$, while the LHC
bounds on the charm quark are considerably weaker that those on the
bottom quark.

\bigskip

\begin{figure}[p]
    % if you want to shrink the plot do it in python not here
	\includegraphics[]{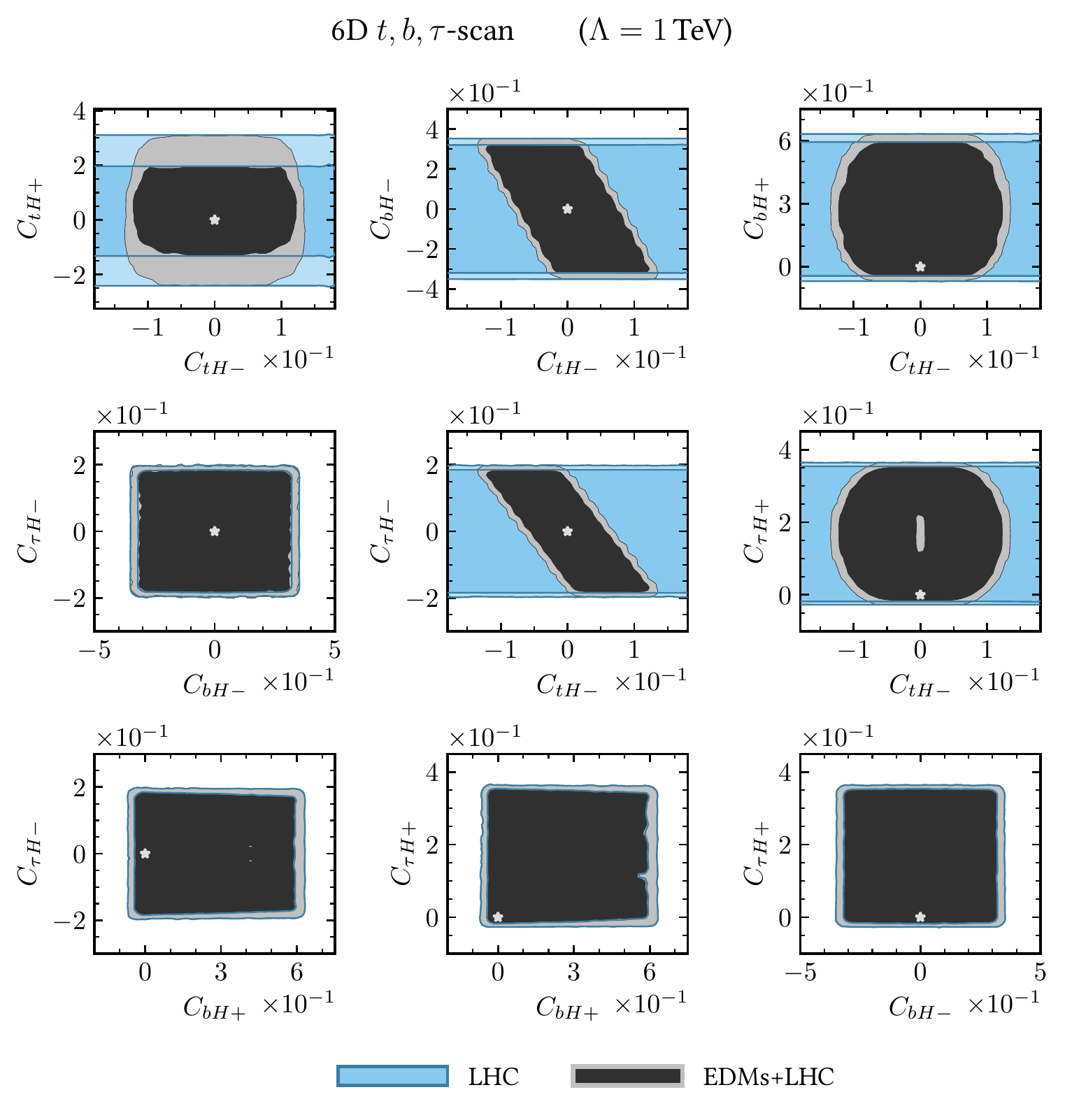}
    \caption{
      Constraints resulting from a 6D scan of top, bottom, and $\tau$ Wilson 
      coefficients  ($C_{tH\pm}$, $C_{bH\pm}$, $C_{\tau H\pm}$) assuming $\Lambda=1$\,TeV.
      In each plot only two parameters are shown, the remaining ones
      are profiled over (see main text).
      Contours represent the allowed 68\% and 95\% confidence regions,
      the colour coding of individual constraints is given in the legend, and 
      gray/black areas correspond to the combined regions.
      For details see main text and the caption of Fig.~\ref{fig:1flavour}.
      \label{fig:tbtau6dscan1}}
\end{figure}

$\bullet$ {\bf Top, bottom, and tau} (third generation)\\
\noindent
In Fig.~\ref{fig:tbtau6dscan1} we present a scan of all six
third-generation parameters $C_{bH\pm}$, $C_{\tau H\pm}$, and
$C_{tH\pm}$.  It is interesting to compare the results to the case in
which only two out of the three flavours were included in the fit
(Fig.~\ref{fig:tbtauscan}).

First, we focus on the set of the four panels (upper and middle row,
center and right) that show the same parameter combinations as the
panels at the same positions in Fig.~\ref{fig:tbtauscan}. We see that
after profiling the remaining third-generation couplings, an
``indirect'' electron EDM constraint on $C_{tH-}$ remains, although
the allowed region is now significantly larger.
By contrast, the ``indirect'' electron EDM constraint can be lifted
completely by $C_{tH-}$ (that is only weakly constrained from LHC
measurements) in the panels in the bottom row and center left of
Fig.~\ref{fig:tbtau6dscan1}, leaving only the LHC constraints. This
should be compared to the corresponding much smaller combined regions
in Fig.~\ref{fig:btauscan}.

The top left panel of Fig.~\ref{fig:tbtau6dscan1} shows the
constraints on both top Wilson coefficients, $C_{tH\pm}$.  Notice that
the allowed values increased by about a factor of two for $C_{tH+}$
and by one order of magnitude for $C_{tH-}$ compared to the
single-flavour scan (Fig.~\ref{fig:1flavour_3rdgen}).

\bigskip

The large computational resources required for scans with more than
six parameters prevent us from scanning over more than three flavours.
Nevertheless, we can infer some results in this direction from the
one-, two-, and three-flavour scans above. Specifically, we will
consider how the constraints on the third-generation Wilson
coefficients change when including more flavours in the scans.

\bigskip

\begin{figure}[t]
    % if you want to shrink the plot do it in python not here
	\includegraphics[]{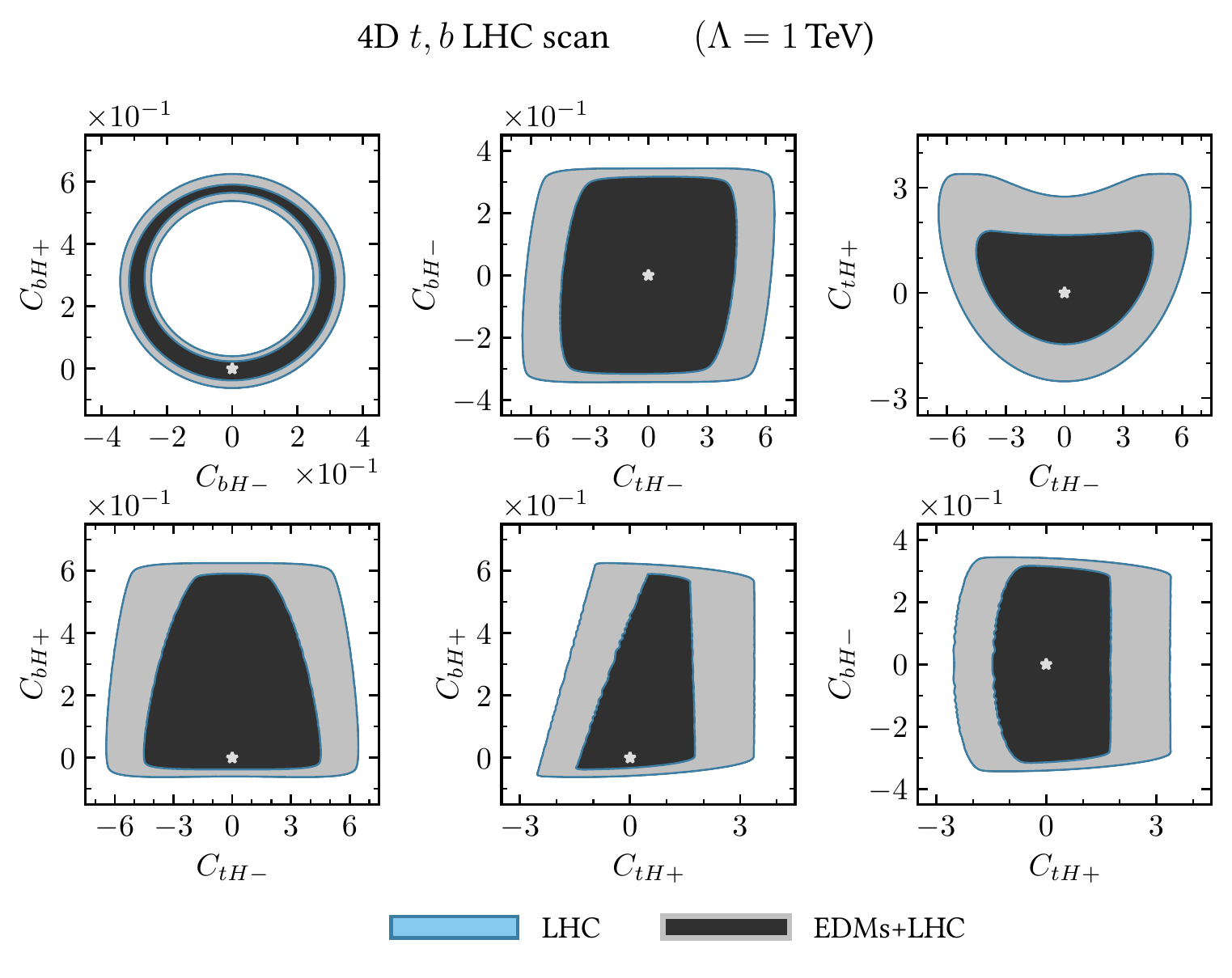}
	\caption{
      LHC constraints resulting from a 4D scan of top and bottom Wilson 
      coefficients  ($C_{tH\pm}$, $C_{bH\pm}$) assuming $\Lambda=1$\,TeV.
      In each plot only two parameters are shown, the remaining ones
      are profiled over (see main text).
      Contours represent the allowed 68\% and 95\% confidence regions,
      the colour coding of individual constraints is given in the legend, and 
      gray/black areas correspond to the combined regions.
      For details see main text and the caption of Fig.~\ref{fig:1flavour}.
      \label{fig:tbscanLHC}}
\end{figure}

\begin{figure}[t]
    % if you want to shrink the plot do it in python not here
	\includegraphics[]{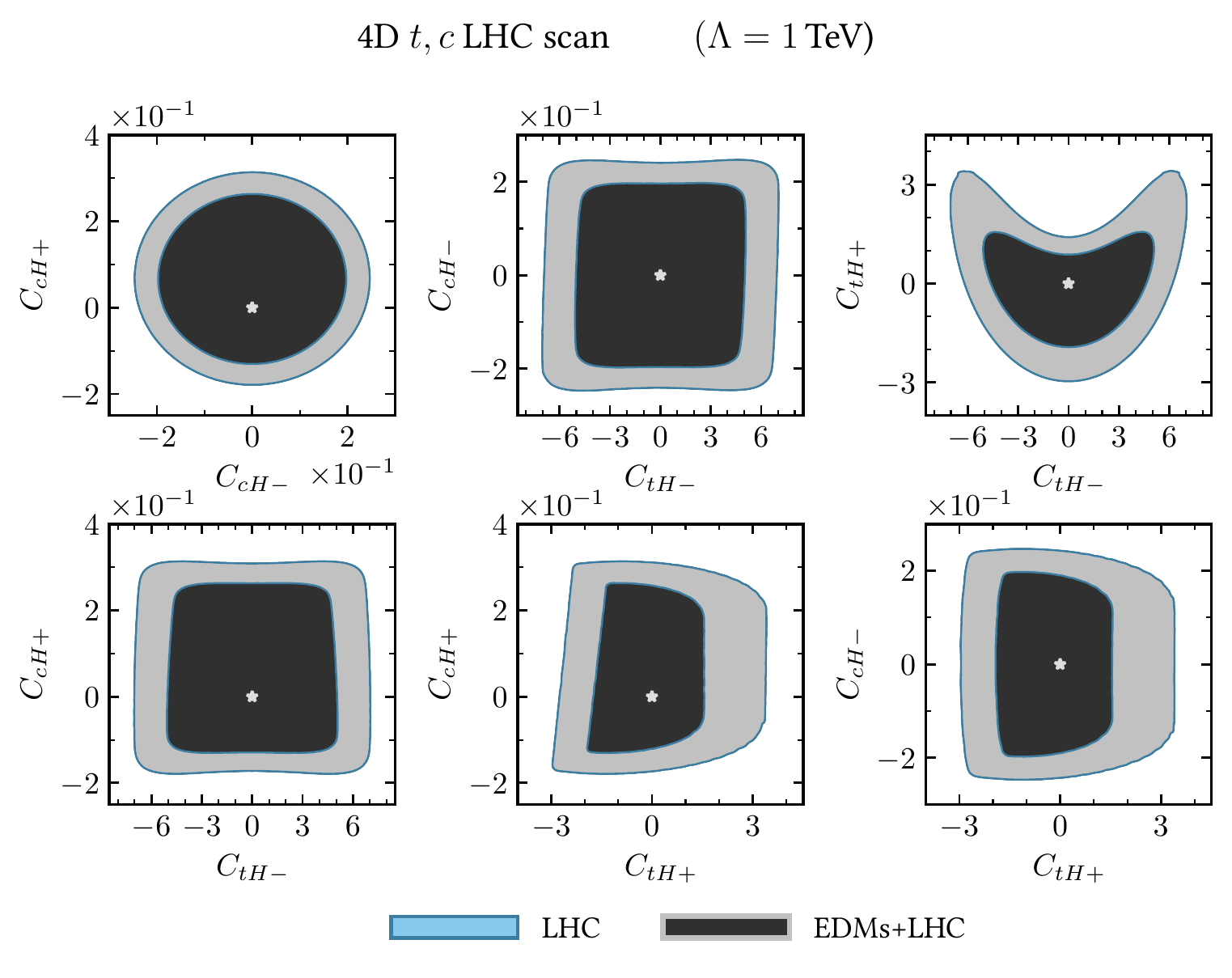}
	\caption{LHC constraints resulting from a 4D scan of top and
          charm Wilson coefficients ($C_{tH\pm}$, $C_{cH\pm}$)
          assuming $\Lambda=1$\,TeV.  In each plot only two parameters
          are shown, the remaining ones are profiled over (see main
          text).  Contours represent the allowed 68\% and 95\%
          confidence regions, the colour coding of individual
          constraints is given in the legend, and gray/black areas
          correspond to the combined regions.  For details see main
          text and the caption of Fig.~\ref{fig:1flavour}.
  \label{fig:tcscanLHC}}
\end{figure}

$\bullet$ {\bf Top, bottom, (electron, light quark)}\\
\noindent
As an example, we float $C_{tH\pm}$ and $C_{bH\pm}$ under the
presumption that any contributions to EDMs can be compensated by
appropriate values of the coefficients $C_{eH-}$ and $C_{qH-}$,
$q=u,d$, which we, however, do not scan over. In this way only LHC
bounds remain, which are one to two orders of magnitude weaker for
$C_{tH-}$ with respect to the electron EDM.  The results are shown in
Fig.~\ref{fig:tbscanLHC}.  We find that the LHC constraints on
$C_{tH+}$ and $C_{bH+}$ are somewhat lifted compared to the
single-flavour scans due to the large contribution of the bottom to
the total Higgs width, while the absence of EDM bounds allows for much
larger allowed ranges of $C_{tH-}$ and $C_{bH-}$. The analogous
results for the charm instead of the bottom couplings are shown in
Fig.~\ref{fig:tcscanLHC}.

\bigskip

More generally, the CP-violating up, down, and electron Wilson
coefficients are merely and severely constrained by EDM measurements.
Hence, including $C_{eH-}$, $C_{uH-}$, and $C_{dH-}$ in a scan over
the Wilson coefficients of the second or third generation would allow
to completely cancel any EDM constraints and again only LHC
constraints would remain.

\subsection{Theory uncertainties\label{sec:had:error}}

In this work, we did not study the impact of theoretical uncertainties
on the bounds on the Wilson coefficients. Hence, a short discussion of
these effects is in order. The relevant uncertainties are: (i)
uncertainties in the hadronic matrix elements; and (ii) perturbative
uncertainties.

(i) The uncertainties on the hadronic matrix elements have been shown
in Sec.~\ref{sec:constraints:EDM}. Note that, in addition to the
ranges given for the parameters, also some of the relative signs are
not determined. In our numerical analysis, we have taken the central
values for the hadronic matrix elements, and (somewhat arbitrarily)
chosen the positive signs where they were not determined. We did not
include these uncertainties in our likelihood function. In fact, no
bounds at the $68\%\,$CL would result from the mercury EDM, while the
neutron EDM bounds would get weaker by about a factor $2$. At
$95\%\,$CL, there would also be no bound from the neutron EDM. (In
Ref.~\cite{Chien:2015xha}, much smaller effects of the hadronic
uncertainties have been found. This may be related to the statistical
treatment of the uncertainties; in our computation we take uncertainty
on the EDMs from hadronic input, $\sigma^2_\text{had}$, to scale with
the Wilson coefficients.)  On the other hand, the bounds on the heavy
quarks (top, bottom, charm) are dominated by the electron EDM that has
no hadronic uncertainties. Recall also, as discussed above, that EDM
constraints become less relevant upon including more Wilson
coefficients in the fit, implying that the hadronic uncertainties do
not (currently) play a significant role in the multi-parameter fits.

(ii) It is important to recognize that, in many cases, the
perturbative uncertainties are as large as the nonperturbative
uncertainties. As an example, consider the bounds on the bottom and
charm Wilson coefficients, studied (within the $\kappa$ framework) in
Ref.~\cite{Brod:2018pli}. There it was shown that the QCD corrections
are large; after inclusion of the two-loop leading-logarithmic QCD
corrections, the uncertainties on the electric and chomoelectric
Wilson coefficients are reduced to order of $30\%$. We do not include
the NLO corrections here, as lattice results are not available for all
required matrix elements (see Refs.~\cite{Bhattacharya:2015rsa,
  Bhattacharya:2016rrc, Yoon:2020soi} for preliminary results). Maybe
somewhat surprisingly, also the electron EDM bound on the CP-violating
bottom and charm Yukawas receives large QCD corrections. The
calculation of these effects is ongoing~\cite{BS2022} and also not
included in this work.
No theory uncertainties are included in the LHC constraints. Note that
they are partially contained in the uncertainties quoted by the
experiments.

\section{Discussion and conclusions\label{sec:conclusions}}

In this work we have presented the first high-dimensional fits of
the coefficients of Yukawa-type SMEFT operators to multiple EDMs and 
LHC data. As expected, upon inclusion of a sufficient number of Wilson
coefficients, all EDM constraints can be evaded, and only LHC bounds
remain. However, when considering the heavy fermions only, a
nontrivial interplay between contributions remains.

There are several ways to further extend our analysis in the future. As
discussed in Sec.~\ref{sec:had:error}, we have neglect the impact of
hadronic and perturbative uncertainties on our fit results. This
impact can be profound in the lower-dimensional scans, and further 
motivates the ongoing efforts to decrease the uncertainties.

Improved experimental bounds (or discoveries), as well as the
inclusion of additional EDMs such as those of proton and deuterium,
once they become available, are expected to have a significant impact
on the global fit. In particular, the different isospin dependence of
the various hadronic systems will help to disentangle CP-odd Higgs
couplings of the first-generation quarks. The CP structure of the
heavy-fermion Yukawas can be tested more directly at present and
future colliders by studying observables that are designed
specifically to test the CP-odd couplings and typically require a vast
amount of (expected) future data (see, e.g.,
Refs.~\cite{Harnik:2013aja, Demartin:2014fia, Buckley:2015vsa,
  Gritsan:2016hjl, Azevedo:2017qiz, Grojean:2020ech,
  Martini:2021uey}).

Finally, as explained in Sec.~\ref{sec:SMEFT} where we introduced the
theoretical framework for flavour-diagonal CP-violation in Higgs
Yukawas, we do in general expect new flavour violating sources to
accompany beyond-the-SM CP-violation.  Therefore, allowing for
flavour-changing contributions within SMEFT, and including the
corresponding observables would be a further extension of the current
analysis.  While such contributions are generically expected in
realistic UV models and are expected to lead to much stronger bounds,
the question is whether, given the proliferation of parameters and
observables, it is not better to study selected UV models directly.

In the final stages of this work, Ref.~\cite{Bahl:2022yrs} was
published, presenting analyses with a similar scope as ours. We
briefly comment on the differences between the two papers. In
Ref.~\cite{Bahl:2022yrs}, the $\kappa$ framework (discussed in this
paper at the end of Sec.~\ref{sec:SMEFT}) is employed to perform
multi-parameter fits to LHC data, including, in particular, the CMS CP
analysis of the $h \to \tau\tau$ decay~\cite{CMS:2021sdq}. The
interplay of LHC data and the electron EDM bound is discussed, but no
combined fit to both LHC and EDM data is performed. By contrast,
constraints that arise from considerations of electroweak baryogenesis
are discussed extensively in Ref.~\cite{Bahl:2022yrs}.
The expression for the electron EDM (Eq.~(13) in
Ref.~\cite{Bahl:2022yrs}) is given in numerical form and thus hard to
verify. However, it seems that the gauge-boson contribution must be
incorrect (either gauge-dependent or containing an implicit
logarithmic dependence on the UV cutoff that is not made explicit; the
cited references contain contradictory results). See our discussion in
Sec.~\ref{sec:match}.

\section*{Acknowledgments}

J.~B. acknowledges support in part by DoE grants DE-SC0020047 and
DE-SC0011784. 
E.~S. would like to thank Luca Merlo and Robert Ziegler for helpful discussions.
 
\appendix

\section{The $\boldsymbol{R_{\xi}}$-gauge Lagrangian\label{sec:full:lagrangian}}

In Eq.~\eqref{eq:Lag:unit:rot} we presented the Yukawa Lagrangian that includes
SMEFT modification from Yukawa operators in the mass-eigenstate basis and in unitarity gauge.
In our computation we work in generic $R_{\xi}$ gauge.
In what follows we present the relevant part of the Lagrangian,
$\Lag_{\text{Yukawa}}$, in $R_{\xi}$ gauge after rotating to the
mass-eigenstate basis for the fermions. 
We split the Lagrangian into
\begin{equation}
  \Lag_{\text{Yukawa}} = \sum_{f=u,d,e}(\Lag_{\text{dim-}4,f} + \Lag_{\text{dim-}5,f} + \Lag_{\text{dim-}6,f}) \,,
\end{equation}
where
\begin{align}
\begin{split}
\Lag_{\text{dim-}4,u} =
  &- m_{u} \bar{u} {u}
   - \bar{u} \left( \frac{m_{u}}{v}      - \frac{v^2}{\sqrt{2}\Lambda^2} C_{{u}H+}   \right)  {u} h\\
  & + \bar{u} \left( \frac{m_{u}}{v} G_0  + \frac{v^2}{\sqrt{2}\Lambda^2} C_{{u}H-} h \right) i \gamma_5 {u}
    + \sqrt{2}\left( G^-\bar{d}_{L} V^\dagger \frac{m_{u}}{v} {u}_R +\text{h.c.}\right) \,,
  \end{split}\\[1em]
\begin{split}
\Lag_{\text{dim-}5,u} =
  & + \frac{v}{2\sqrt{2}\Lambda^2}\bar{u}\left( C_{{u}H+}(3h^2+ G_0^2 +2 G^+G^-) +2 C_{{u}H-} G_0 h \right) {u} \\
  & + \frac{v}{2\sqrt{2}\Lambda^2}\bar{u}\left( C_{{u}H-}(3h^2+ G_0^2 +2 G^+G^-) -2 C_{{u}H+} h G_0  \right) i \gamma_5 {u} \\
  & - \frac{v}{\Lambda^2}                \left( \bar{d}_L V^\dagger(C_{{u}H+}+i C_{{u}H-})  {u}_R h G^- +\text{h.c.}\right)\,,
\end{split}\\[1em]
\begin{split}
\Lag_{\text{dim-}6,u} =
  & + \frac{1}{2\sqrt{2}\Lambda^2}(h^2+ G_0^2+ 2G^+ G^-)\bar{u}\left( C_{{u}H+}h +C_{{u}H-} G_0\right) {u} \\
  & + \frac{1}{2\sqrt{2}\Lambda^2}(h^2+ G_0^2+ 2G^+ G^-)\bar{u}\left( C_{{u}H-}h -C_{{u}H+} G_0\right) i \gamma_5 {u} \\
  & - \frac{1}{2\Lambda^2}(h^2+ G_0^2+ 2G^+ G^-)               \left( \bar{d}_L V^\dagger(C_{{u}H+}+i C_{{u}H-}) {u}_R G^-+\text{h.c.}\right) \,.
\end{split}
\end{align}
Here, $u = (u,c,t)$, $d = (d,s,b)$, $m_u = (m_u, m_c, m_t)$, and the $C_{uH\pm}$ can also contain flavour off-diagonal pieces.
The corresponding Lagrangian for down-type quarks is obtained by the
obvious interchanges $u\leftrightarrow d$, $G^+\leftrightarrow G^-$,
$V^\dagger \leftrightarrow V$, and by a reversal of sign in all terms
that are odd in the Goldstone fields.  Analogously, the corresponding
Lagrangian for leptons is obtained from the up-type Lagrangian above
by the obvious interchanges $u\rightarrow e$, $d\rightarrow \nu$,
$G^+\leftrightarrow G^-$, $V^\dagger \rightarrow 1$, and again by
reversing the sign in all terms that are odd in the Goldstone fields.

\section{An alternative flavour basis: real and diagonal Yukawas\label{sec:nir:basis}}

There is a certain freedom of choosing the flavour basis for the SMEFT
Lagrangian to use for presenting the phenomenological constraints.
This freedom stems from the fact that both the dimension-four Yukawas
($Y_f$) and the dimension-six SMEFT coefficients ($C'_{fH}$)
contribute to the observed masses (and mixings) of fermions.  As long
as this is guaranteed any basis is equivalent.  However, since a UV
extension of the SM matched to SMEFT would in general induce
contributions both to $Y_f$ and $C'_{fH}$ there is no clear notion of
a ``better'' or a more ``physical'' basis for $Y_f$. We have thus
opted to present the bounds in the mass-eigenstate basis as discussed
in Sec.~\ref{sec:SMEFT} (see also
Ref.~\cite{Alonso-Gonzalez:2021jsa}). The advantage of this basis is
that it is the one required for computations and is also directly
related to the $\kappa$-framework basis (see discussion around
Eq.~\eqref{eq:kappa}).  Another approach would be to attempt to
provide the constraints in terms of basis-independent, i.e.,
Jarlskog-type, invariants (see recent Ref.~\cite{Bonnefoy:2021tbt}).

In this appendix, we discuss a different choice of basis that has been
used in the literature \cite{Fuchs:2020uoc}.  We comment on the
differences and point out a consistency condition on the Wilson
coefficients in this basis that has been missed in the literature.  As
before, our starting point is the fermion mass term in
Eq.~\eqref{eq:fermion:mass:term}:
\begin{equation}
  \Lag_{\text{mass}} = - \sum_{f=u,d,\ell} \frac{v}{\sqrt{2}} \bar{f}_L
  \left(Y_f - \frac{v^2}{2\Lambda^2} C_{fH}' \right)f_R
  +\text{h.c.}\,,
\end{equation}
with $Y_f$ and $C'_{fH}$ generic, complex $3\times 3$ matrices.
Now, instead of diagonalising the full matrix in parentheses, as we
did, one may choose to rotate the fermion fields by a biunitary transformation
that diagonalises only the dimension-four Yukawa matrices $Y_f$. 
In other words, we rotate with the transformation matrices
\begin{equation}\label{eq:UW:rotation:nir}
  f_L \to \hat{U}_f f_L\,, \qquad f_R \to \hat{W}_f f_R \,,
\end{equation}
for $f = u,d,\ell$, such that after the rotation
\begin{equation}
  \Lag_{\text{mass}} = - \sum_{u,d,\ell}
  \frac{v}{\sqrt{2}} \bar{f}_L \left(\hat{Y}_f - \frac{v}{2\Lambda^2} \hat{C}_{fH}\right) f_R
  + \text{h.c.} \,,
\end{equation}
where now the matrices $\hat{Y}_f \equiv \hat{U}_f^\dagger Y_f
\hat{W}_f$ are diagonal and real with entries that, however, {\itshape
  do not} correspond to the observed fermion masses. The kinetic terms
of quarks is also affected by the transformation in
Eq.~\eqref{eq:UW:rotation:nir}. At this stage this means that there is
a matrix $\hat{V}\equiv U_u^\dagger U_d$ in the kinetic
terms. Moreover, we have defined $\hat{C}_{fH} \equiv
\hat{U}_f^\dagger C_{fH} \hat{W}_f$. In general, the matrices
$\hat{C}_{fH}$ are not diagonal.  However, similarly to the discussion
of Sec.~\ref{sec:SMEFT}, one can make the UV assumption that they turn
out to be complex but diagonal (or ignore off-diagonal entries).  This
is possible if the matrices $U_f$, $W_f$ simultaneously diagonalise
both $Y_f$ and $C_{fH}$.  Below we assume that this is the case.  The
analysis of Ref.~\cite{Fuchs:2020uoc} uses this basis, i.e.,
$\hat{C}_{fH\pm}$ or rescalings of them, to present the
phenomenological constraints on the SMEFT parameter space.

Since in general, the elements of $\hat{C}_{fH}$ contain phases, we still
have not rotated to the mass-eigenstate basis. To this end we perform
an additional (flavour-diagonal) rotation on the right-handed fields 
as in Ref.~\cite{Fuchs:2020uoc}:
\begin{equation}
  f_{R,i} \rightarrow e^{i\theta_{f,i}} f_{R,i} \,, \quad f = u,d,\ell \,.
\end{equation}
Note that this chiral rotation leaves the kinetic term invariant. The
phases $\theta_{f,i}$ are fixed such that they absorb any phase in the
corresponding $\hat{C}_{fH}$. Therefore after the rotation
\begin{equation}
  \Lag_{\text{mass}} = - \sum_{u,d,\ell}
  \frac{v}{\sqrt{2}} \bar{f}_L y_f^{\text{SM}} f_R
  +\text{h.c.}\,.
\end{equation}
Here, the $y_f^{\text{SM}}$ are diagonal and real matrices with
entries that correspond to the observed fermion masses, i.e., we have
$m_f = \frac{v}{\sqrt{2}} y_f^{\text{SM}}$, obtained from
\begin{equation}
y^{\text{SM}}_f
\equiv
\left( \hat{Y}_f - \frac{v^2}{2\Lambda^2} \hat C_{fH}\right) e^{i\theta_f}
=
\hat{U}_f^\dagger \left(Y_f - \frac{v^2}{2\Lambda^2} C_{fH}\right) \hat{W}_f e^{i\theta_f} \,.
\end{equation}
The phases $\theta_f$ are thus fixed and can be computed as functions
of the physical masses and the $\hat{C}_{fH}$ entries via
\begin{equation}
  \sin\theta_{f,i}
= \frac{v}{2\sqrt{2} m_{f,i}} \frac{v^2}{\Lambda^2} \hat{C}_{fH-}
= \frac{v}{2\sqrt{2} m_{f,i}} \frac{v^2}{\Lambda^2} {\rm Im}[\hat{U}_f^\dagger C_{fH} \hat{W}_f] \,.
\end{equation}
As we have the restriction $|\sin\theta_{f,i}| \leq 1$, this gives a
consistency bound on $\hat{C}_{fH-}$ or equivalently on 
${\rm Im}[\hat{U}_f^\dagger C_{fH} \hat{W}_f]$.

\addcontentsline{toc}{section}{References}
\bibliographystyle{JHEP}
\bibliography{references}

\end{document}